# The New Generalized Odd Median Based Unit Rayleigh with a New Shape Oscillating Hazard Rate Function

Iman M. Attia *

[Imanattiathesis1972@gmail.com](mailto:Imanattiathesis1972@gmail.com) ,[imanattia1972@gmail.com](mailto:imanattia1972@gmail.com)

*Department of Mathematical Statistics, Faculty of Graduate Studies for Statistical Research, Cairo University, Egypt*

## Abstract:

In this paper, the author presents the generalized form of the Median-Based Unit Rayleigh (MBUR) distribution, a novel statistical distribution that is specifically defined within the interval (0, 1) expressing oscillating hazard rate function. This generalization adds a new parameter to the MBUR distribution that significantly addresses the unique characteristics of data represented as ratios and proportions, which are commonly encountered in various fields of research. The establishment of this generalization aims to deepen our understanding of these phenomena by providing a robust framework for analysis. The paper offers a thorough and meticulous derivation of the probability density function (PDF) for the MBUR distribution, illuminating each phase of the process with clarity and precision. It delves deep into the intricacies of the MBUR distribution's properties, presenting a rigorous examination of the accompanying functions that are vital for robust statistical evaluation. These functions—comprising the cumulative distribution function (CDF), survival function, hazard rate and reversed hazard rate function. The paper discusses real data analysis and how the generalization improves such analysis.

# Introduction

Fitting data to statistical distributions is crucial for understanding the underlying processes that generate the data. Researchers have successfully developed a variety of distributions to describe complex real-world phenomena effectively. Before 1980, the primary techniques for generating distributions included solving systems of differential equations, using transformations, and applying quantile function strategies. Since 1980, the focus has shifted to adding new parameters to existing distributions or combining known distributions. These methods have resulted in a wide range of adaptable and flexible distributions that can accommodate various types of asymmetrical data and outliers in data sets. Fitting distributions to data improves modeling in analyses related to regression, survival analysis, reliability analysis, and time series analysis.

A multitude of real-world phenomena can be elegantly captured as proportions, ratios, or fractions nestled within the bounded interval (0,1). These captivating representations are not merely abstract concepts; they reflect the intricate relationships found in various fields such as biology, where the delicate balance of ecosystems is analyzed; finance, where the ebb and flow of market ratios unfold; and mortality rates, which provide profound insights into human health and longevity. Additionally, recovery rates in medical science showcase the resilience of life, while economics delves into the nuanced distributions of wealth and resources. Engineering and hydrology further enrich this tapestry, modeling everything from structural integrity to the flow of water in our environments. The measurement sciences, too, rely on these continuous distributions, breathing life into data that inform our understanding of the world.

Some of these distributions include: the Johnson SB distribution (1), Beta distribution (2), Unit Johnson distribution (3), Topp-Leone distribution (4), Unit Gamma distribution (5),(6),(7),(8), Unit Logistic distribution (9), Kumaraswamy distribution (10), Unit Burr-III distribution (11), Unit Modified Burr-III distribution (12), Unit Burr-XII distribution (13), Unit-Gompertz distribution (14), Unit-Lindley distribution (15), Unit-Weibull distribution (16), Unit-Birnbaum-Saunders (17) and Unit Muth distribution (18).

This paper is structured into several sections for clarity and coherence. Section 1 provides a comprehensive discussion of the methodology employed to derive the new distribution. Section 2 delves into its fundamental characteristics, including the probability density function (PDF), cumulative distribution function (CDF), survival function (S), hazard function (HF), reversed hazard function (RHF), and quantile function. Section 3 offers an in-depth discussion that encompasses an analysis of real data as well as a detailed examination of the findings. In conclusion, Section 4 provides a comprehensive overview of our findings and offers valuable recommendations for future research, inviting further exploration and innovation in the field.

# Section 1: Methodology:

## Derivation of the PDF

The general formula of the median order statistics for an odd sample size can be written as in equation (1)

$$f_{i:m} = \frac{m!}{\left(\frac{m-1}{2}\right)!\left(\frac{m-1}{2}\right)!}[F_x(x)]^{\frac{m-1}{2}}[1-F_x(x)]^{\frac{m-1}{2}}f_x(x) \ldots \ldots \ldots \ldots \ldots \ldots \ldots (1)$$

Where (i) is the i<sup>th</sup> odd order statistics and m is the sample size. Replacing (m) sample size (which is an odd number) with m=2n+1 as shown in equation (2.A) & (2.B)

$$f_{i:m} = \frac{(2n+1)!}{\left(\frac{2n+1-1}{2}\right)!\left(\frac{2n+1-1}{2}\right)!}[F_x(x)]^{\frac{2n+1-1}{2}}[1-F_x(x)]^{\frac{2n+1-1}{2}}f_x(x) \ldots \ldots \ldots \ldots (2.A)$$

$$f_{i:m} = \frac{(2n+1)!}{(n)!\,(n)!}[F_x(x)]^n[1-F_x(x)]^n f_x(x) \ldots \ldots \ldots \ldots \ldots \ldots \ldots \ldots \ldots (2.B)$$

Substituting equation 3 which is the CDF and the PDF of Rayleigh distribution in equation 2.B yields equation 4.

$$F_x(x) = 1 - exp\left(\frac{-x^2}{\alpha^2}\right) \ \& \ f_x(x) = \frac{2x}{\alpha^2}exp\left(\frac{-x^2}{\alpha^2}\right) \ldots \ldots \ldots \ldots \ldots \ldots \ldots \ldots (3)$$

$$f_{i:m} = \frac{(2n+1)!}{(n)!\,(n)!}\left[1 - exp\left(\frac{-x^2}{\alpha^2}\right)\right]^n \left[exp\left(\frac{-x^2}{\alpha^2}\right)\right]^n \frac{2x}{\alpha^2}exp\left(\frac{-x^2}{\alpha^2}\right) \ldots \ldots \ldots \ldots \ldots (4)$$

Using the transformation in equation 5 and the Jacobian in equation 6 and then substituting both in equation 4 yields the new PDF in equation 7. This is the first version of generalization of the MBUR.

$$let\ y = e^{-x^2} \ \ldots \ldots \ldots \ldots \ldots \ldots \ldots \ldots \ldots \ldots \ldots \ldots \ldots \ldots \ldots \ldots \ldots \ldots \ldots \ldots \ldots \ldots (5.A)$$

$$-\ln y = x^2 \ , \ [-\ln y]^{0.5} = x \ \ldots \ldots \ldots \ldots \ldots \ldots \ldots \ldots \ldots \ldots \ldots \ldots \ldots \ldots \ldots \ldots \ldots (5.B)$$

$$\frac{dx}{dy} = \frac{1}{2}[-\ln y]^{-0.5}\frac{-1}{y} \ \ldots \ldots \ldots \ldots \ldots \ldots \ldots \ldots \ldots \ldots \ldots \ldots \ldots \ldots \ldots \ldots \ldots \ldots (6)$$

$$f_y(y) = \frac{\Gamma(2n+2)}{\Gamma(n+1)\Gamma(n+1)}\frac{1}{\alpha^2}\left[1 - y^{\alpha^{-2}}\right]^n [y]^{\frac{n+1}{\alpha^2}-1}, n \geq 0, \alpha > 0, 0 < y < 1 \ \ldots \ldots \ldots \ldots (7)$$

The second version of generalization is obtained by substituting the CDF and the PDF of Rayleigh in equation 8 which yields equation 9

$$f_{i:n} = \frac{n!}{\left(\frac{n-1}{2}\right)!\left(\frac{n-1}{2}\right)!}[F_x(x)]^{\frac{n-1}{2}}[1 - F_x(x)]^{\frac{n-1}{2}} f_x(x) \ldots \ldots \ldots \ldots \ldots \ldots \ldots \ldots \ldots \ldots \ldots (8)$$

$$f_{i:n} = \frac{n!}{\left(\frac{n-1}{2}\right)!\left(\frac{n-1}{2}\right)!}\left[1 - exp\left(\frac{-x^2}{\alpha^2}\right)\right]^{\frac{n-1}{2}}\left[exp\left(\frac{-x^2}{\alpha^2}\right)\right]^{\frac{n-1}{2}} \frac{2x}{\alpha^2} exp\left(\frac{-x^2}{\alpha^2}\right) \ldots \ldots \ldots \ldots (9)$$

Substituting the same transformation of equation 5 and the same Jacobian of equation 6 in equation 9 yields the new PDF in equation 10. This is the second version of generalization of the MBUR distribution.

$$f_y(y) = \frac{\Gamma(n+1)}{\Gamma\left(\frac{n}{2}+\frac{1}{2}\right)\Gamma\left(\frac{n}{2}+\frac{1}{2}\right)} \frac{1}{\alpha^2}[1 - y^{\alpha^{-2}}]^{\frac{n-1}{2}}[y]^{\frac{n+1}{2\alpha^2}-1}, n \geq 1, \alpha > 0, 0 < y < 1 \ldots \ldots (10)$$

**Theorem 1**: Both versions in equation 7 and 10 are valid PDF.

**Proof**: PDF version in equation 7:

To prove that the PDF in equation 7 is a valid PDF, the integral in equation 11 should equal 1. Applying the transformation in equation 12 and substitute in equation 11:

$$\frac{\Gamma(2n+2)}{\Gamma(n+1)\Gamma(n+1)} \int_0^1 \frac{1}{\alpha^2}[1 - y^{\alpha^{-2}}]^n [y]^{\frac{n+1}{\alpha^2}-1} dy = 1 \ldots \ldots \ldots \ldots \ldots \ldots \ldots \ldots \ldots \ldots \ldots (11)$$

$$let: \quad y^{\frac{1}{\alpha^2}} = w, \ so \ y = w^{\alpha^2} \ this \ gives \ dy = \alpha^2 w^{\alpha^2-1} dw \ldots \ldots \ldots \ldots \ldots \ldots \ldots \ldots \ldots (12)$$

$$\frac{\Gamma(2n+2)}{\Gamma(n+1)\Gamma(n+1)} \int_0^1 \frac{1}{\alpha^2}[1 - w]^n [w^{\alpha^2}]^{\frac{n+1}{\alpha^2}-1} \alpha^2 w^{\alpha^2-1} dw = 1$$

For the PDF version in equation 10, applying the same transformation will integrate the PDF in equation 10 to 1.

## Section 2: Some properties of the generalized odd MBUR distribution

**Theorem 2**: the cumulative distribution function (CDF) of the generalized odd MBUR is:

$P(Y < y) = I_w(n+1, n+1)$ for version 1 & $P(Y < y) = I_w\left(\frac{n+1}{2}, \frac{n+1}{2}\right)$ for version 2.

**Proof**: for version 1:

$$P(Y < y) = \frac{\Gamma(2n+2)}{\Gamma(n+1)\Gamma(n+1)} \int_0^y \frac{1}{\alpha^2} \left[1 - y^{\alpha^{-2}}\right]^n [y]^{\frac{n+1}{\alpha^2}-1} dy \quad \ldots \ldots \ldots \ldots \ldots \ldots \ldots (13)$$

Apply the transformation of equation 12 and substitute in equation 13 yields equation 14

$$P(Y < y) = \frac{\Gamma(2n+2)}{\Gamma(n+1)\Gamma(n+1)} \int_0^w [1-w]^n [w]^n \, dw = \frac{B_w(n+1, n+1)}{B(n+1, n+1)} \quad \ldots \ldots \ldots \ldots (14)$$

For version 2:

$$P(Y < y) = \frac{\Gamma(n+1)}{\Gamma\left(\frac{n}{2}+\frac{1}{2}\right)\Gamma\left(\frac{n}{2}+\frac{1}{2}\right)} \int_0^y \frac{1}{\alpha^2} \left[1-y^{\alpha^{-2}}\right]^{\frac{n-1}{2}} [y]^{\frac{n+1}{2\alpha^2}-1} dy \quad \ldots \ldots \ldots \ldots (15)$$

Apply the transformation of equation 12 and substitute in equation 15 yields equation 16

$$P(Y < y) = \frac{\Gamma(n+1)}{\Gamma\frac{(n+1)}{2}\Gamma\left(\frac{n+1}{2}\right)} \int_0^w [1-w]^{\frac{n-1}{2}} [w]^{\frac{n-1}{2}} dw = \frac{B_w\left(\frac{n+1}{2}, \frac{n+1}{2}\right)}{B\left(\frac{n+1}{2}, \frac{n+1}{2}\right)} \quad \ldots \ldots \ldots (16)$$

**Lemma 1**: the survival function (S) for version 1 is shown in equation 17

$$S(y) = 1 - P(Y < y) = 1 - I_w(n+1, n+1) \quad \ldots \ldots \ldots \ldots \ldots \ldots \ldots \ldots \ldots \ldots (17)$$

**Lemma 2**: the survival function (S) for version 2 is shown in equation 18

$$S(y) = 1 - P(Y < y) = 1 - I_w\left(\frac{n+1}{2}, \frac{n+1}{2}\right) \quad \ldots \ldots \ldots \ldots \ldots \ldots \ldots \ldots (18)$$

**Lemma 3**: the Hazard function or rate (HF or hr) for version 1 is shown in equation 19

$$hr(y) = \frac{f_Y(y)}{1 - P(Y < y)} = \frac{\frac{\Gamma(2n+2)}{\Gamma(n+1)\Gamma(n+1)} \frac{1}{\alpha^2} \left[1 - y^{\frac{1}{\alpha^2}}\right]^n [y]^{\frac{n+1}{\alpha^2}-1}}{1 - I_w(n+1, n+1)} \quad \ldots \ldots \ldots \ldots (19)$$

**Lemma 4**: the Hazard function or rate (HF or hr) for version 2 is shown in equation 20

$$hr(y) = \frac{f_Y(y)}{1 - P(Y < y)} = \frac{\frac{\Gamma(n+1)}{\Gamma\left(\frac{n}{2}+\frac{1}{2}\right)\Gamma\left(\frac{n}{2}+\frac{1}{2}\right)} \frac{1}{\alpha^2} \left[1-y^{\alpha^{-2}}\right]^{\frac{n-1}{2}} [y]^{\frac{n+1}{2\alpha^2}-1}}{1 - I_w(n+1, n+1)} \quad \ldots \ldots \ldots (20)$$

**Lemma 5**: the reversed hazard function (RHF) or reversed hazard rate (rhr) for version 1 is shown in equation 21

$$rhr(y) = \frac{f_Y(y)}{P(Y<y)} = \frac{\frac{\Gamma(2n+2)}{\Gamma(n+1)\Gamma(n+1)}\frac{1}{\alpha^2}[1-y^{\alpha^{-2}}]^n [y]^{\frac{n+1}{\alpha^2}-1}}{I_w(n+1,n+1)} \quad\ldots\ldots\ldots\ldots\ldots (21)$$

**Lemma 6**: the reversed hazard function (RHF) or reversed hazard rate (rhr) for version 2 is shown in equation 22

$$rhr(y) = \frac{f_Y(y)}{P(Y<y)} = \frac{\frac{\Gamma(n+1)}{\Gamma\left(\frac{n}{2}+\frac{1}{2}\right)\Gamma\left(\frac{n}{2}+\frac{1}{2}\right)}\frac{1}{\alpha^2}[1-y^{\alpha^{-2}}]^{\frac{n-1}{2}} [y]^{\frac{n+1}{2\alpha^2}-1}}{I_w(n+1,n+1)} \quad\ldots\ldots\ldots (22)$$

The quantile function of the distribution has no closed explicit form.

Figures (1-9) illustrate the PDF of the first version for different values of n {2, 3, 5, 10, 20, 30, 40, 50, 84} and different values of alpha {0.452, 1, 1.5, 1.8, 2.2, 4.5, 10.5}

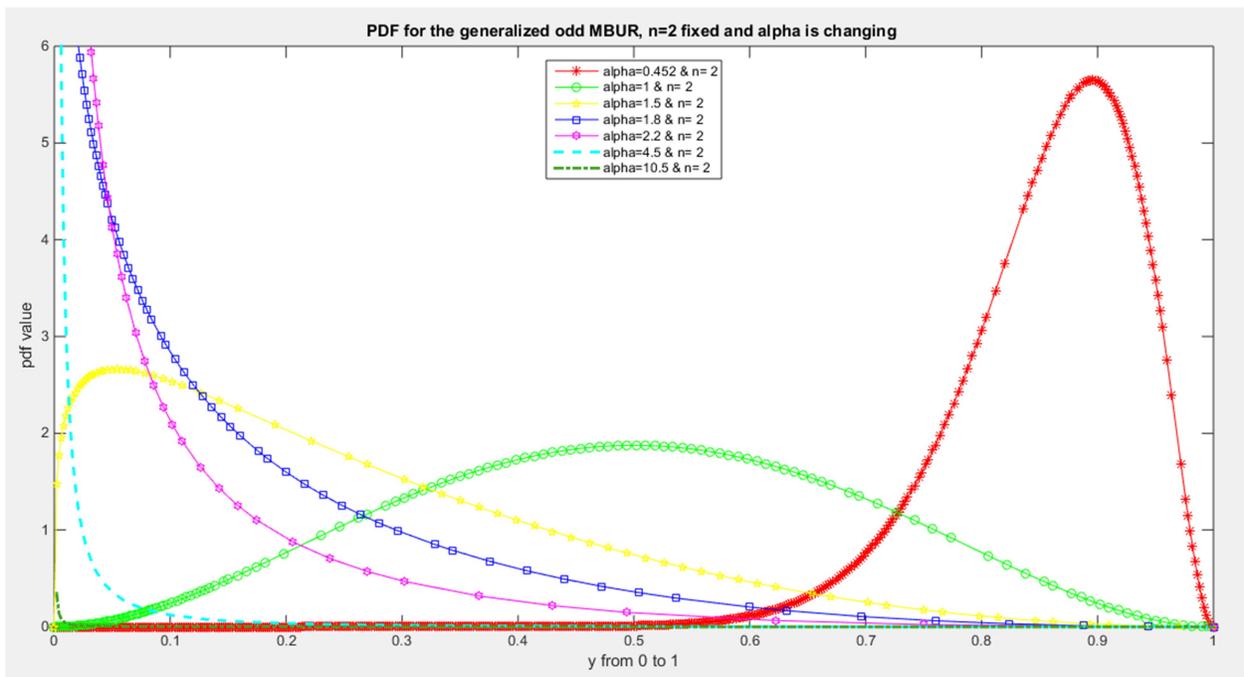

Fig. 1 shows PDF of the first version for different alpha and n=2. For alpha less than 1, the PDF is left skewed. If alpha is one the PDF is symmetric around y=0.5. For alpha values larger than 1 the PDF exhibits right skewness or decreasing .

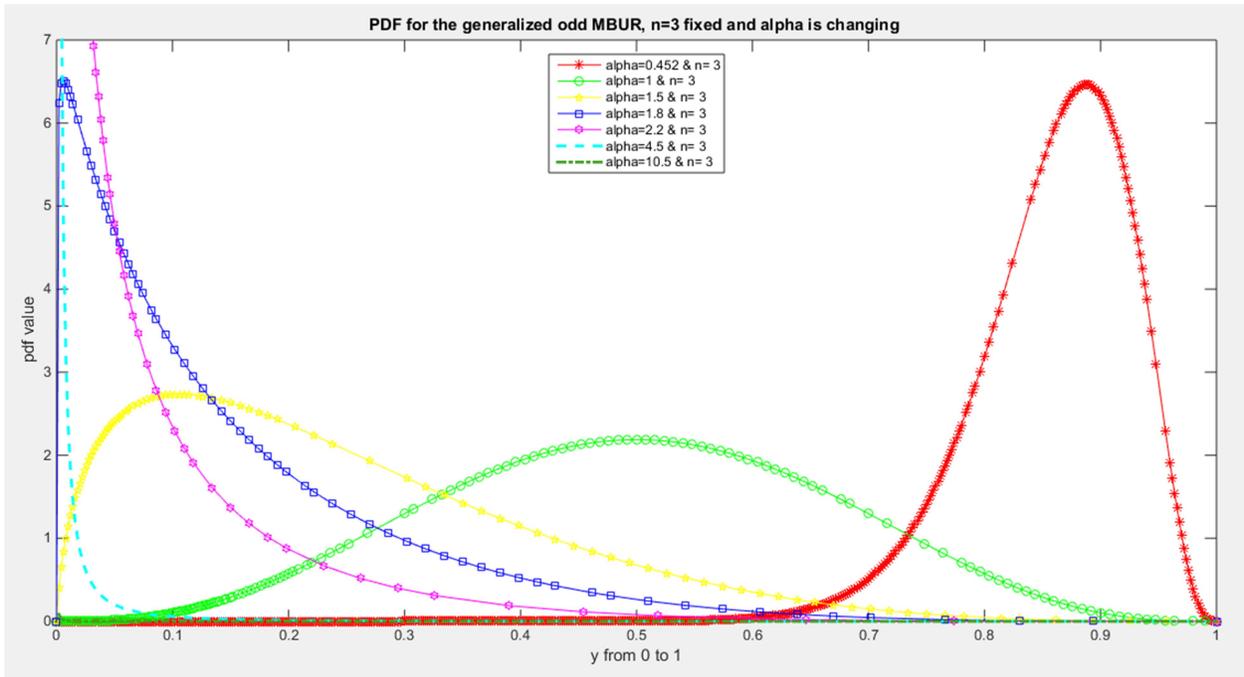

Fig. 2 shows PDF of the first version for different levels of alpha and n=3

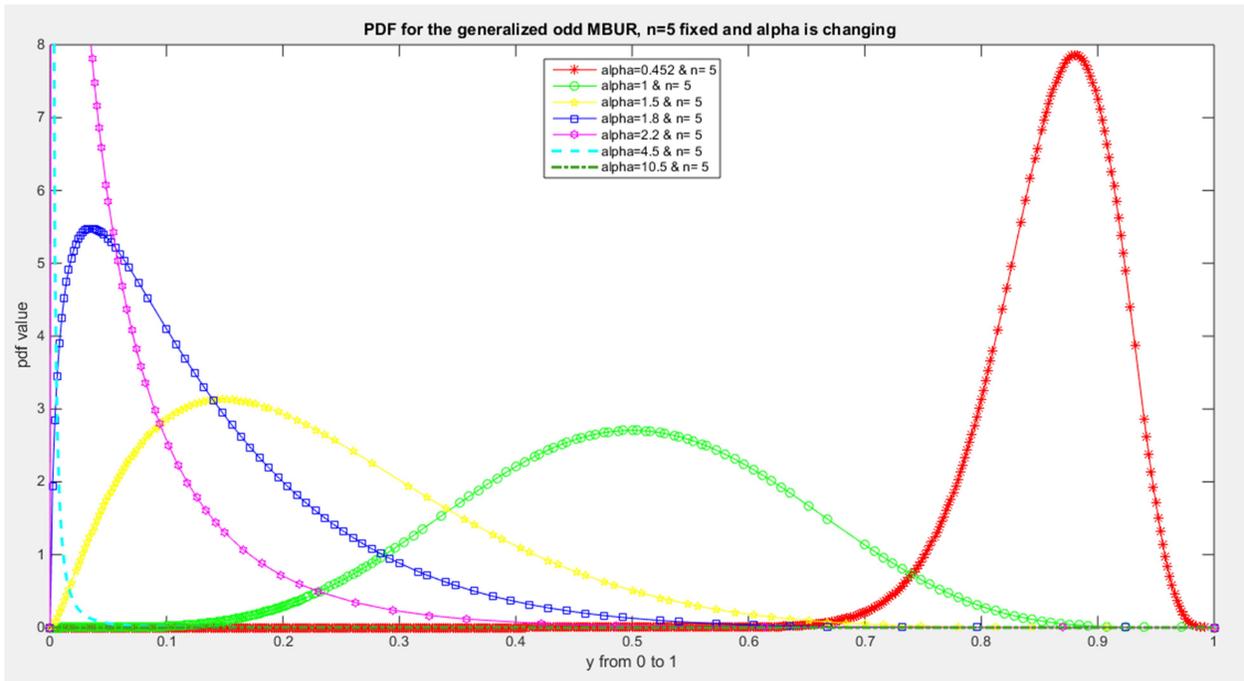

Fig. 3 shows PDF of the first version for different levels of alpha and n=5

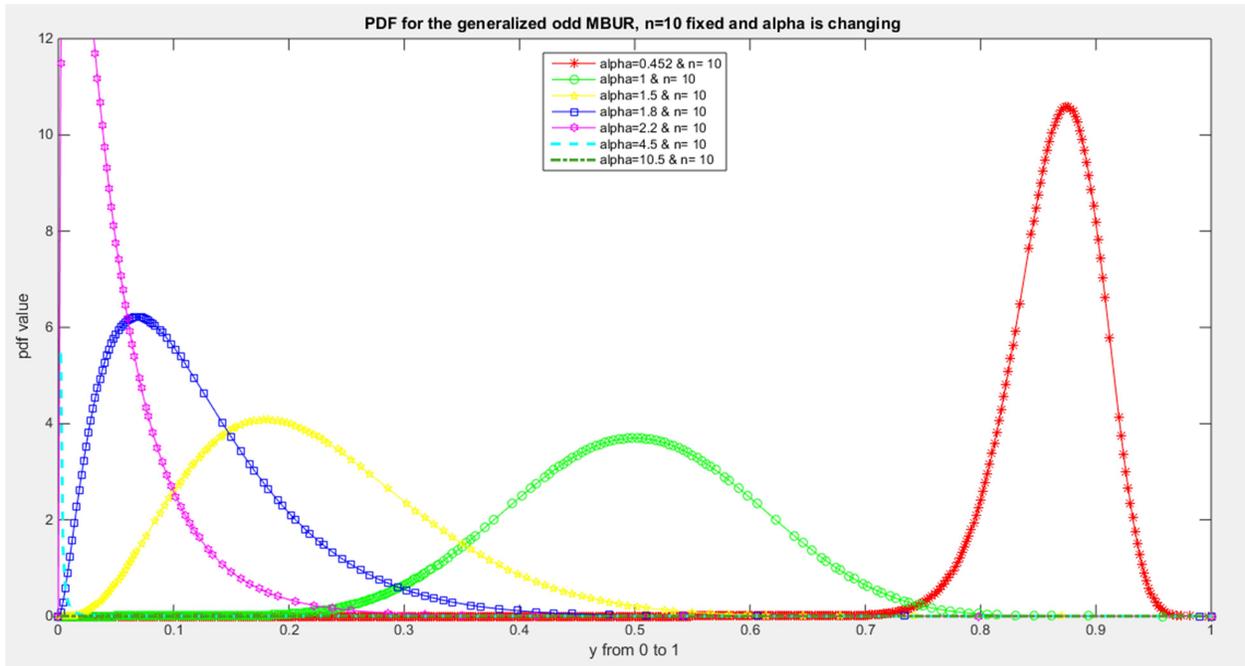

Fig. 4 shows PDF of the first version for different levels of alpha and n=10

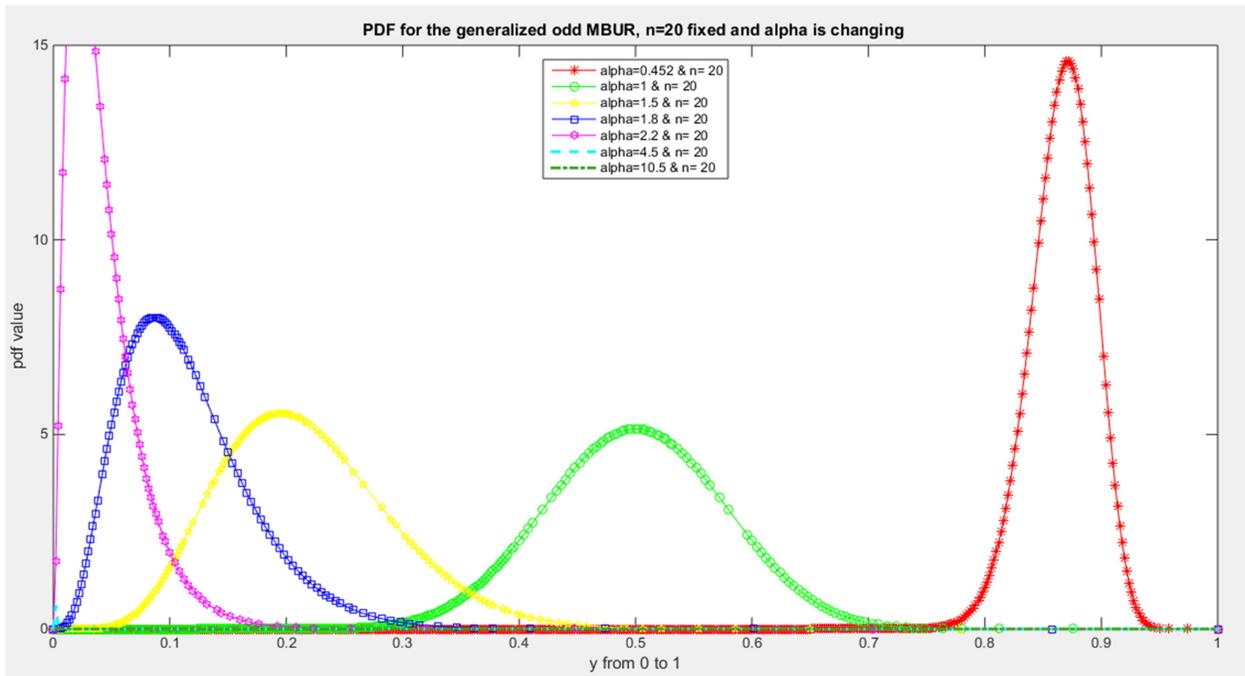

Fig. 5 shows PDF of the first version for different levels of alpha and n=20

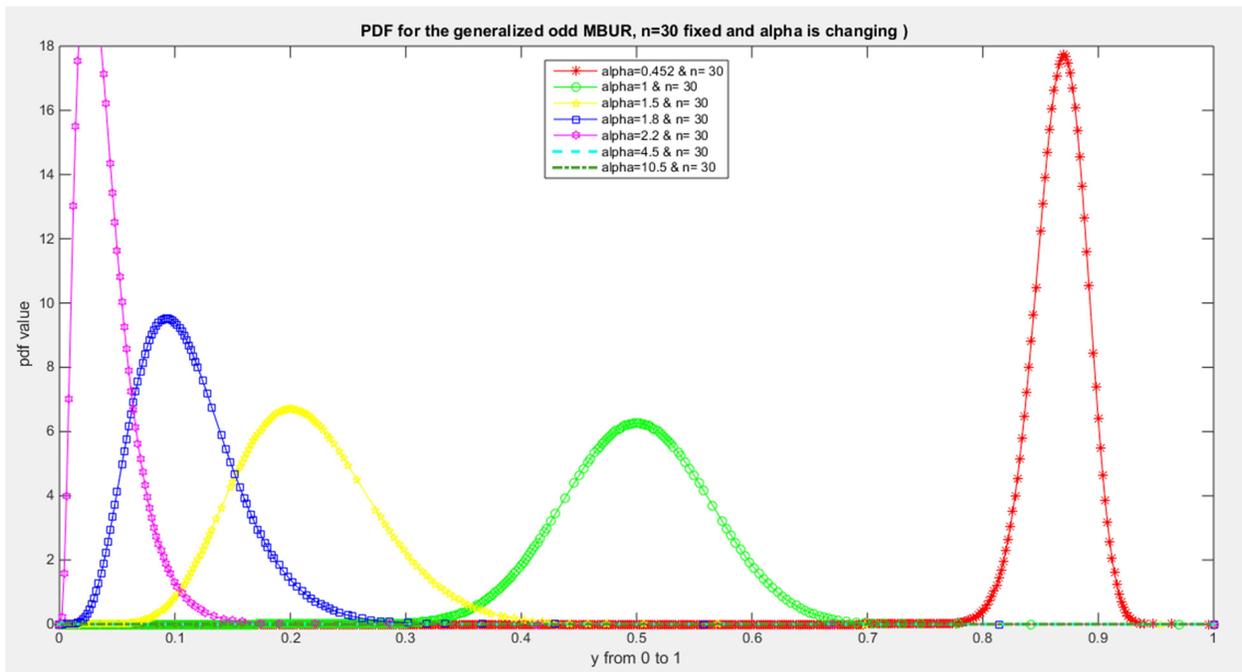

Fig. 6 shows PDF of the first version for different levels of alpha and n=30

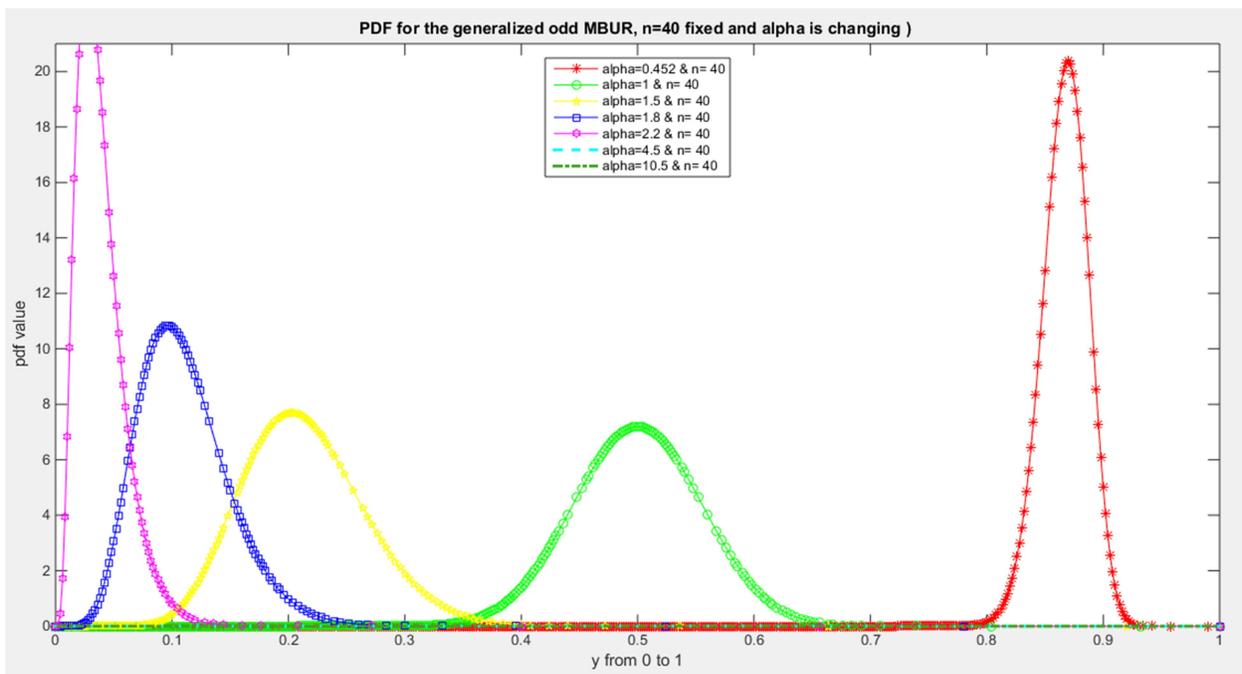

Fig. 7 shows PDF of the first version for different levels of alpha and n=40

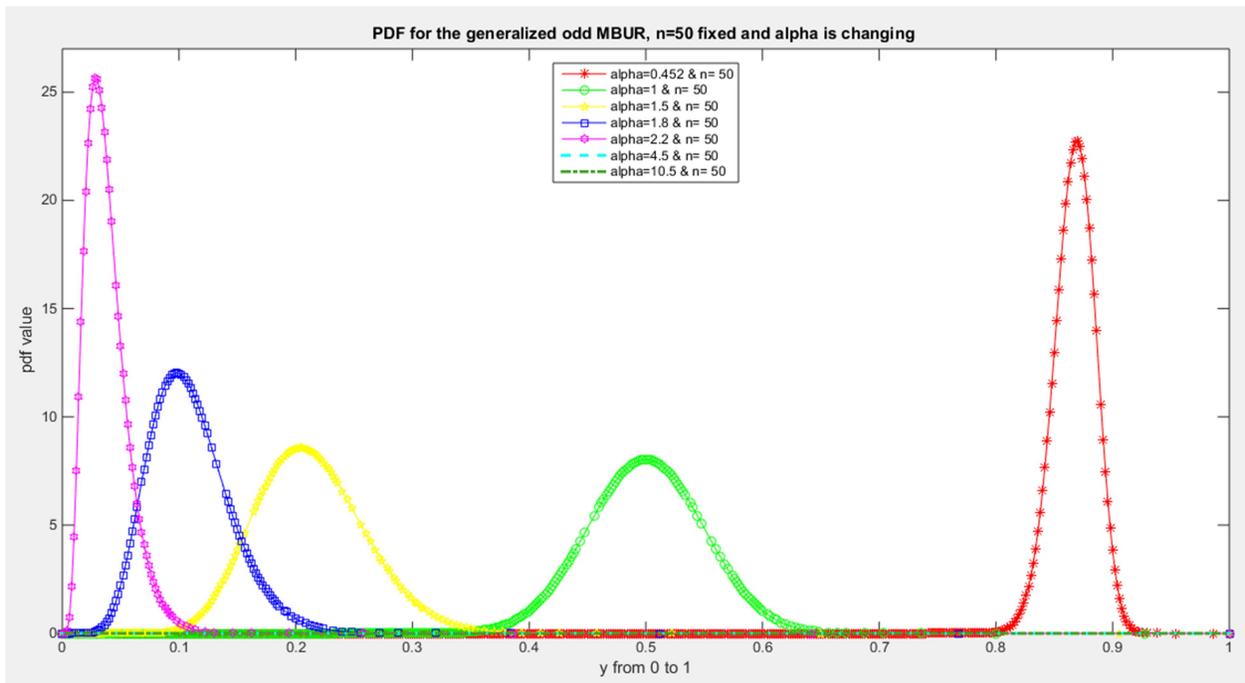

Fig. 8 shows PDF of the first version for different levels of alpha and n=50

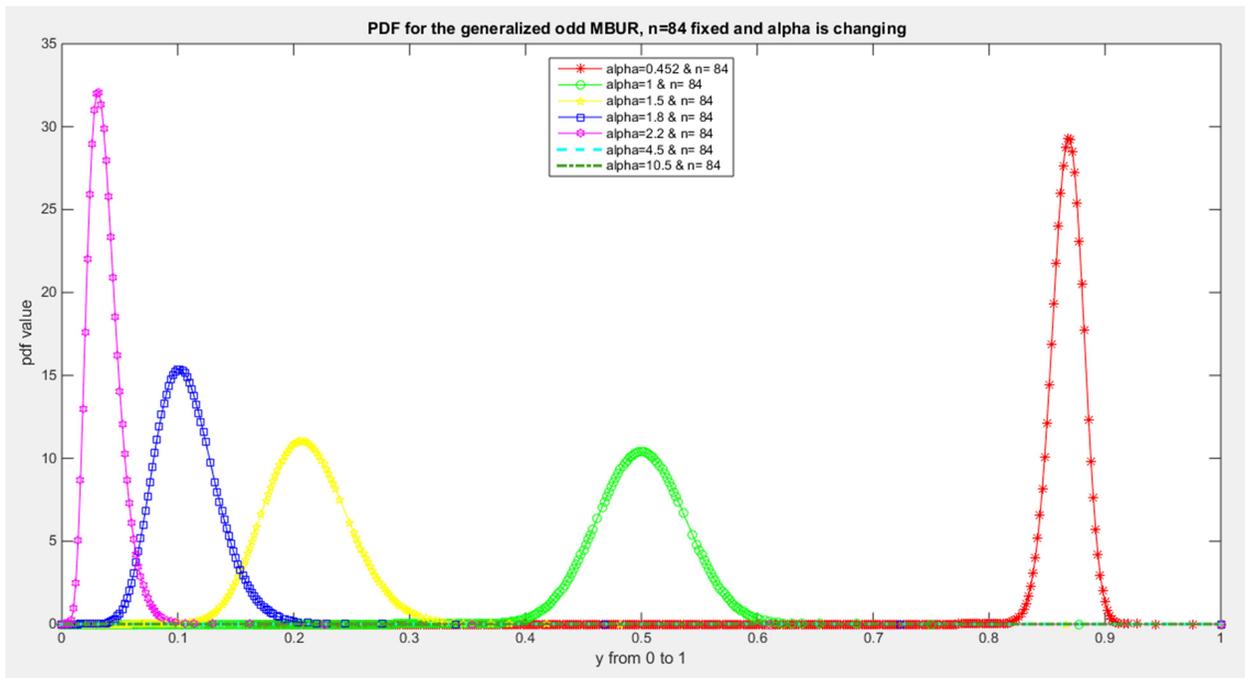

Fig. 9 shows PDF of the first version for different levels of alpha and n=84

The figures of the PDF of the first version illustrates that for values of n larger than 30 but not exceeding 84 and associated with different values of alpha not exceeding 4, these PDFs are more or less symmetric around different values of the variable occupying the whole unit range. This is an advantage of the new added parameter that generalizes the distribution.

Figures (10-20) illustrate the PDF of the second version for different values of n {2, 3, 5, 10, 20, 30, 40, 50, 100, 150, 170} and different values of alpha {0.452, 1, 1.5, 1.8, 2.2, 4.5, 10.5}

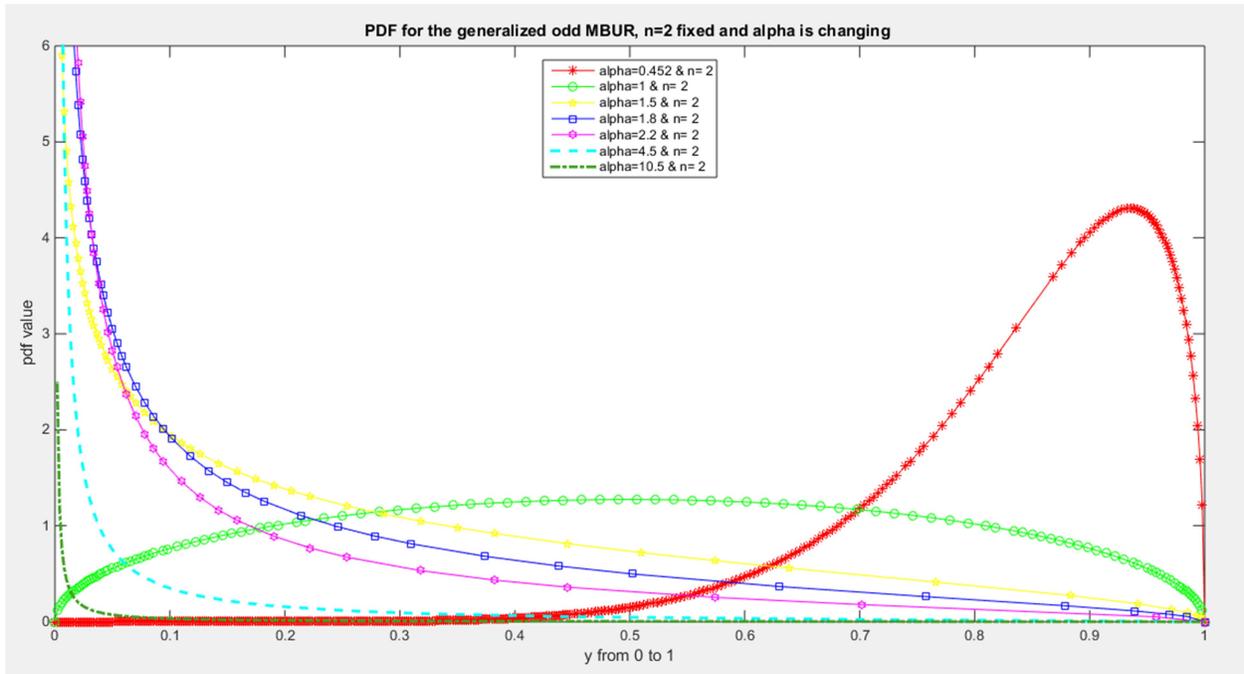

Fig. 10 shows PDF of the second version for different levels of alpha and n=2

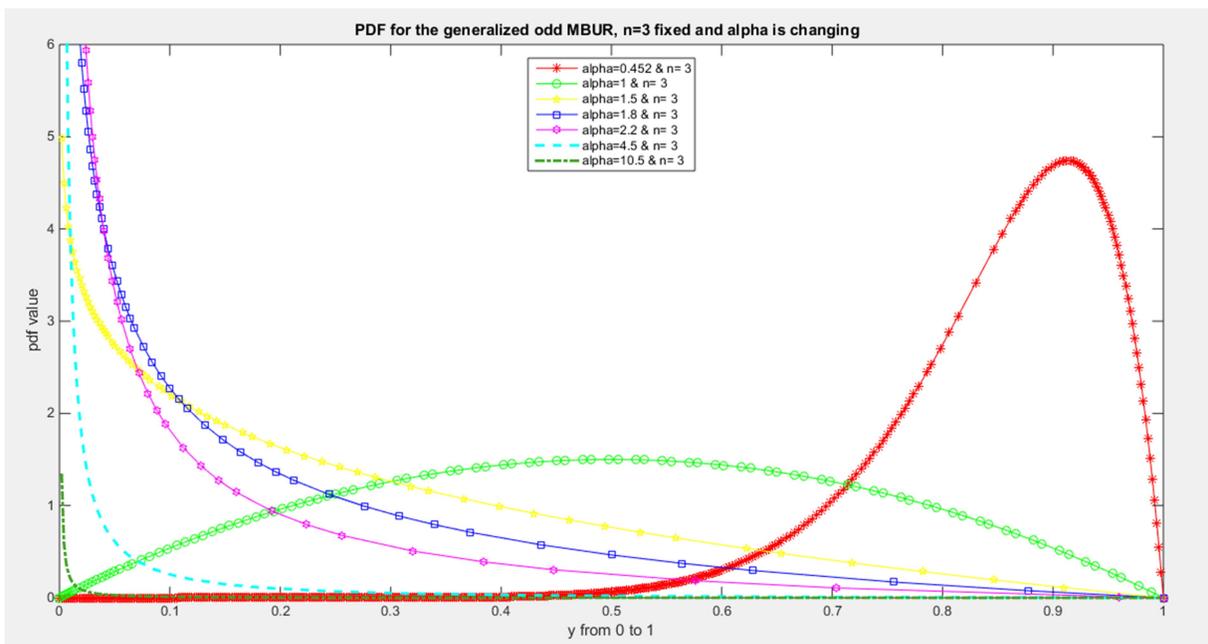

Fig. 11 shows PDF of the second version for different levels of alpha and n=3

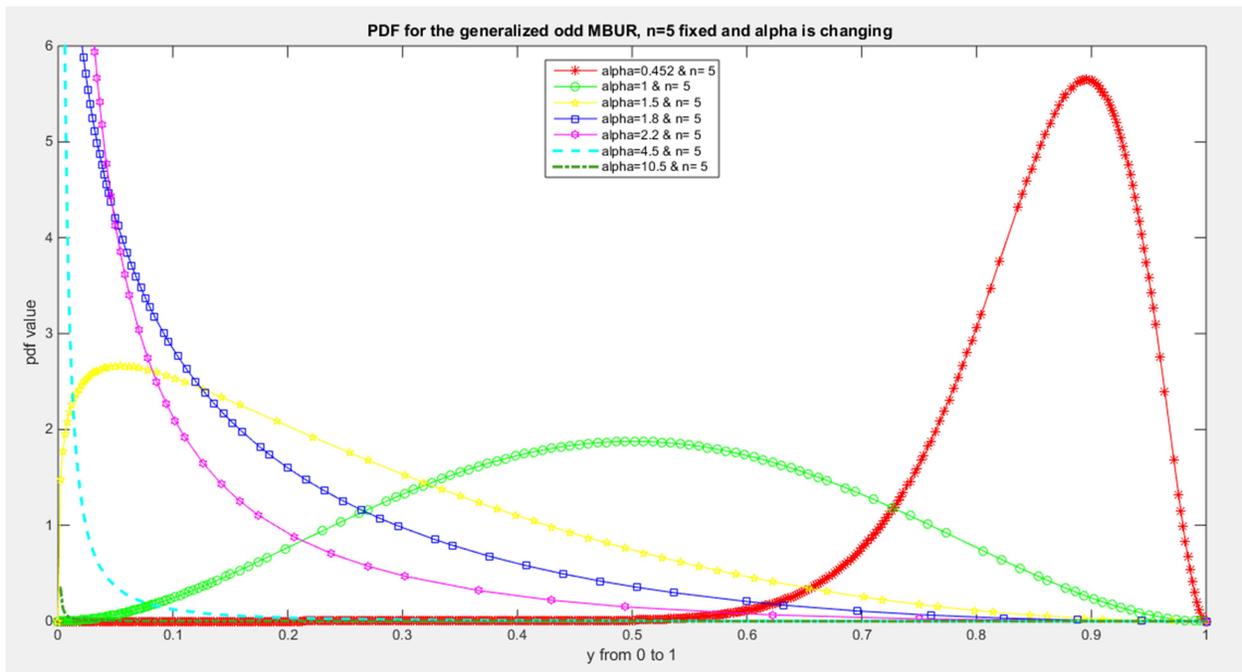

Fig. 12 shows PDF of the second version for different levels of alpha and n=5

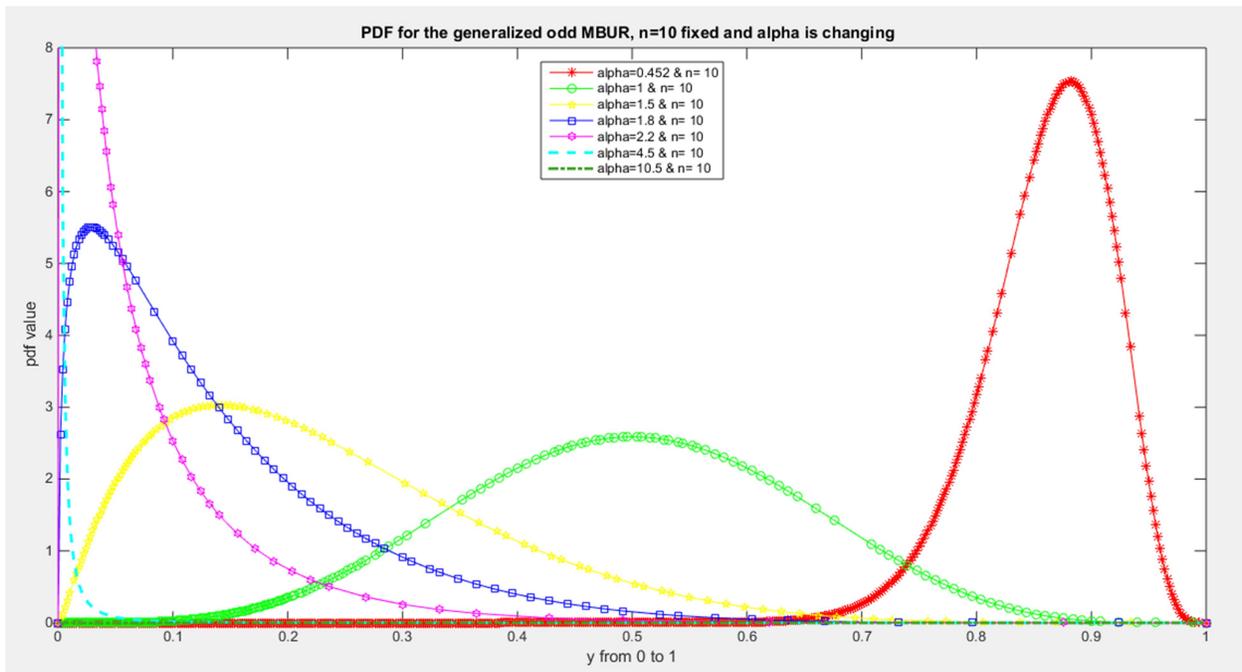

Fig. 13 shows PDF of the second version for different levels of alpha and n=10

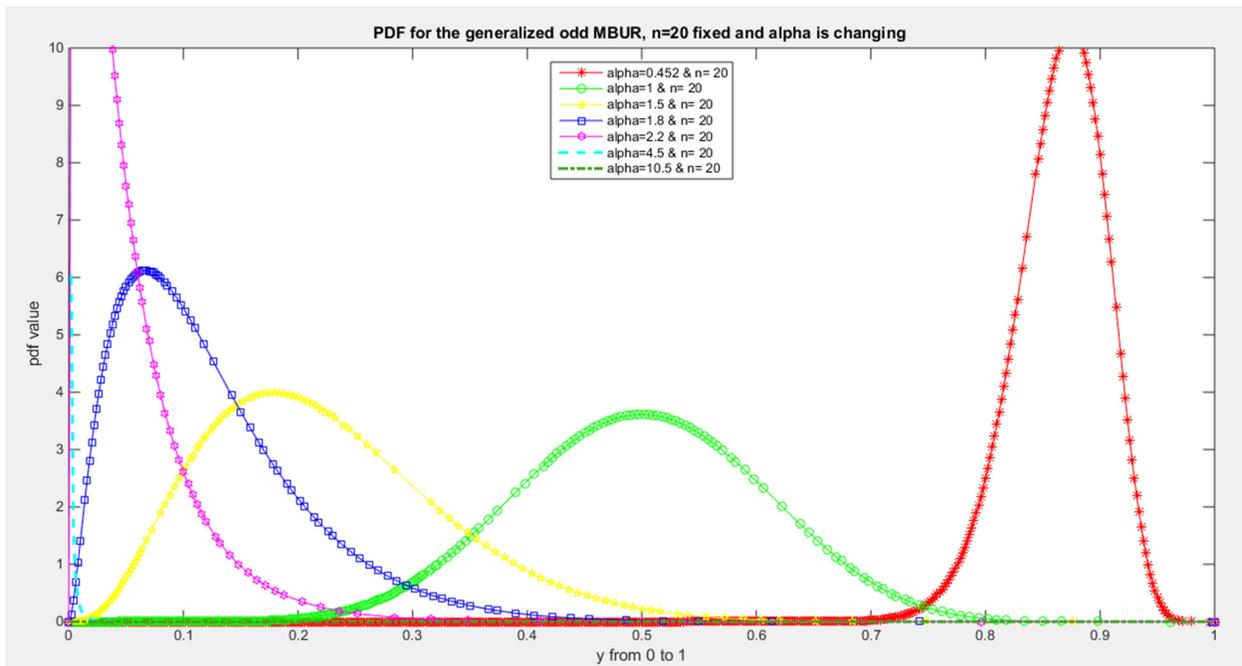

Fig. 14 shows PDF of the second version for different levels of alpha and n=20

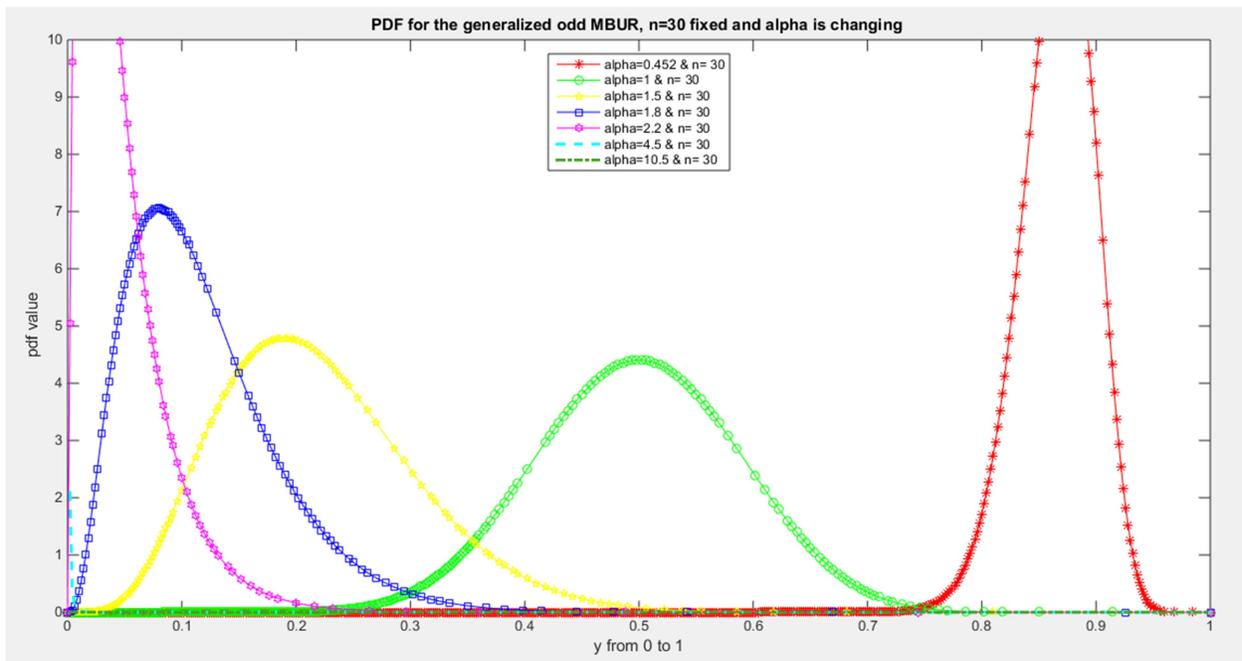

Fig. 15 shows PDF of the second version for different levels of alpha and n=30

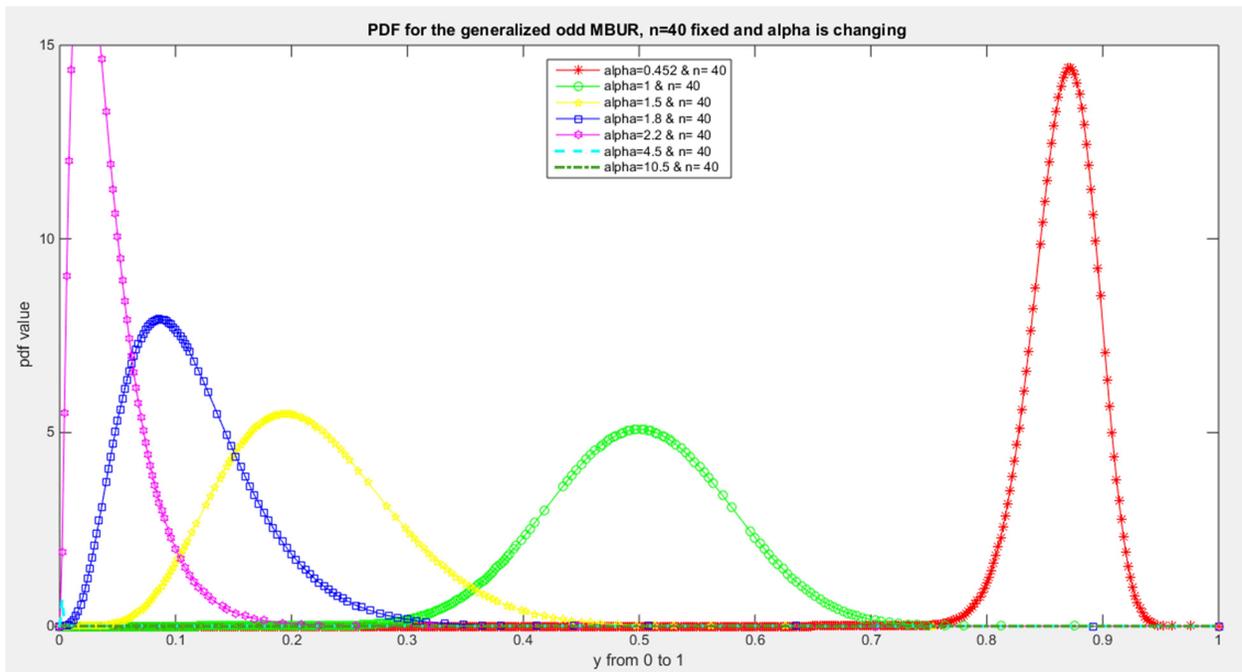

Fig. 15 shows PDF of the second version for different levels of alpha and n=40

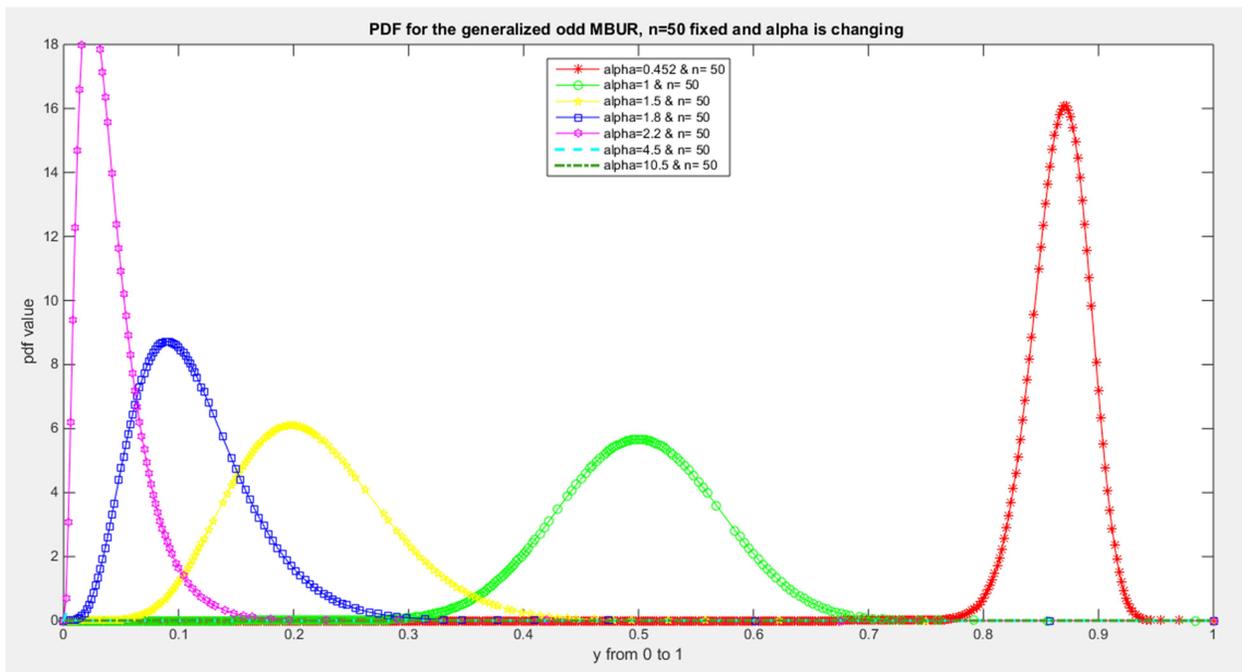

Fig. 16 shows PDF of the second version for different levels of alpha and n=50

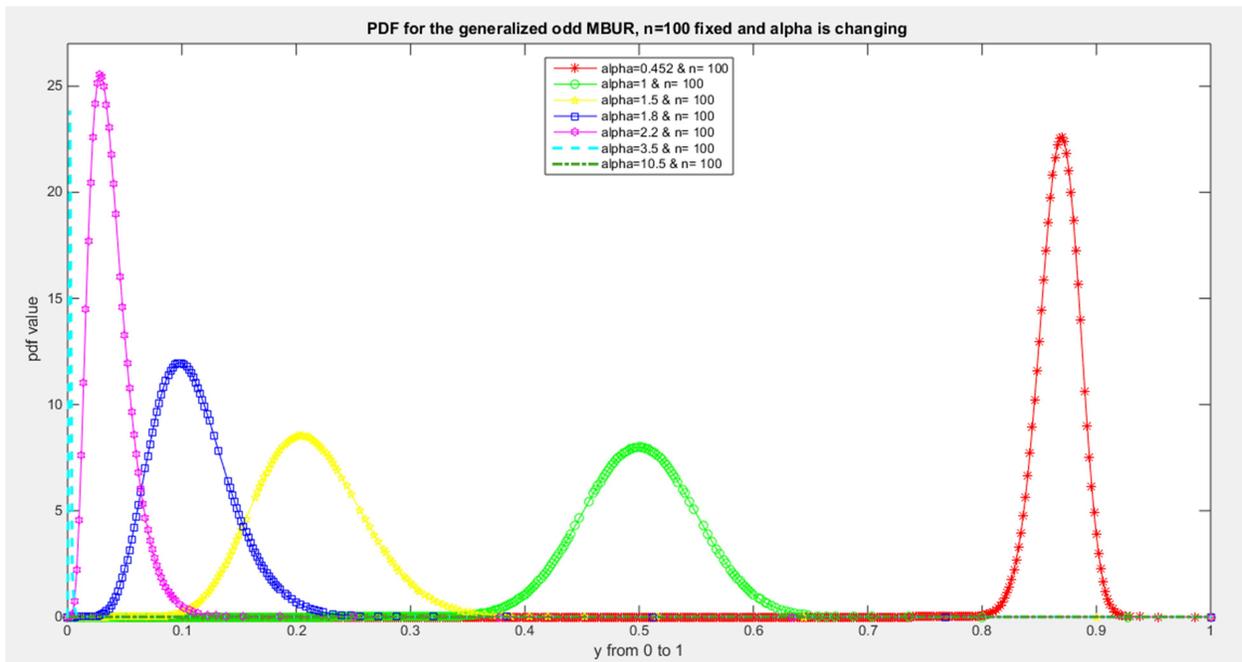

Fig. 17 shows PDF of the second version for different levels of alpha and n=100

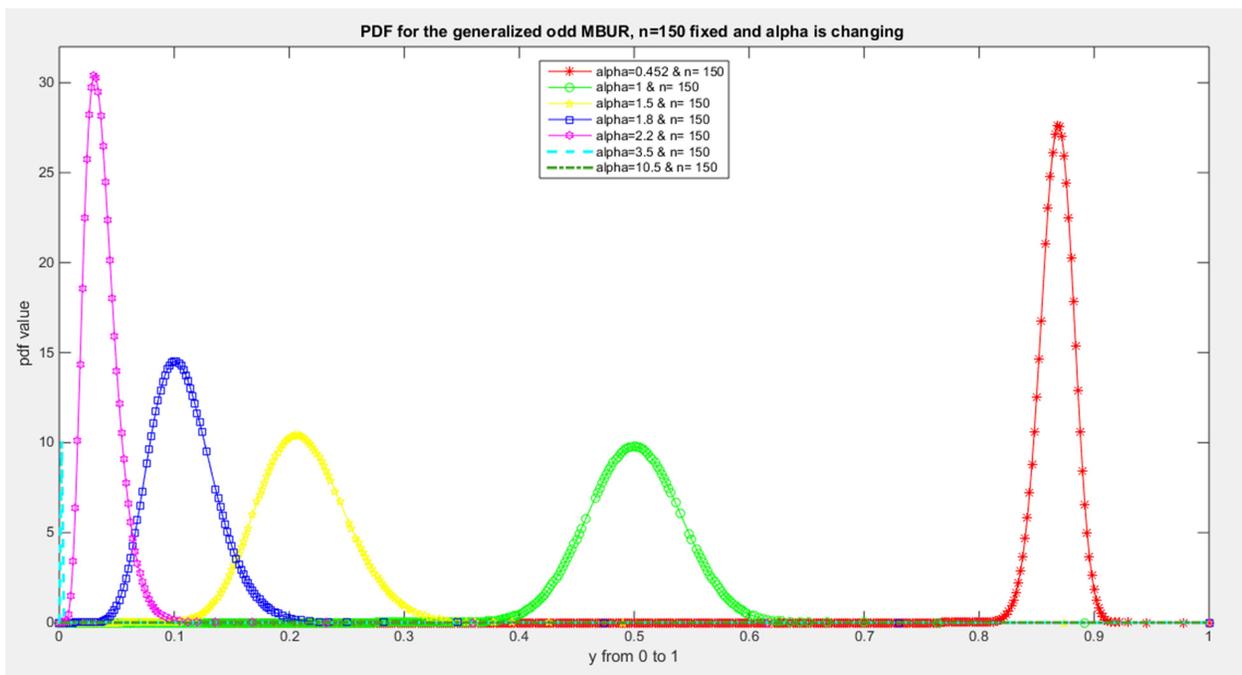

Fig. 18 shows PDF of the second version for different levels of alpha and n=150

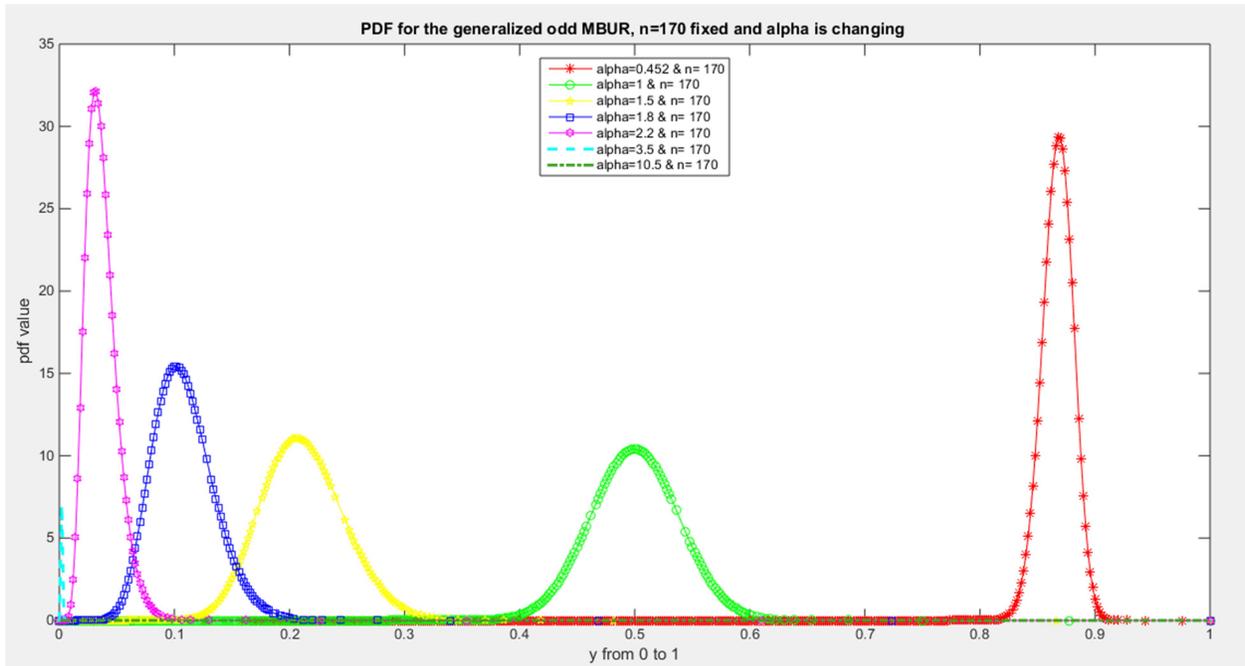

Fig. 19 shows PDF of the second version for different levels of alpha and n=170

The figures of the PDF of the second version illustrates that for values of n larger than 50 but not exceeding 170 and associated with different values of alpha not exceeding 4, these PDFs are more or less symmetric around different values of the variable occupying the whole unit range. This is an advantage of the new added parameter that generalizes the distribution

Figures (20-21) illustrate the CDF of the first version for different values of n {2, 3, 5, 10, 20, 30, 40, 50, 100, 150, 170} and different values of alpha {0.272, 0.614, 1, 1.3, 1.8, 2.2, 3.5, 10.5}

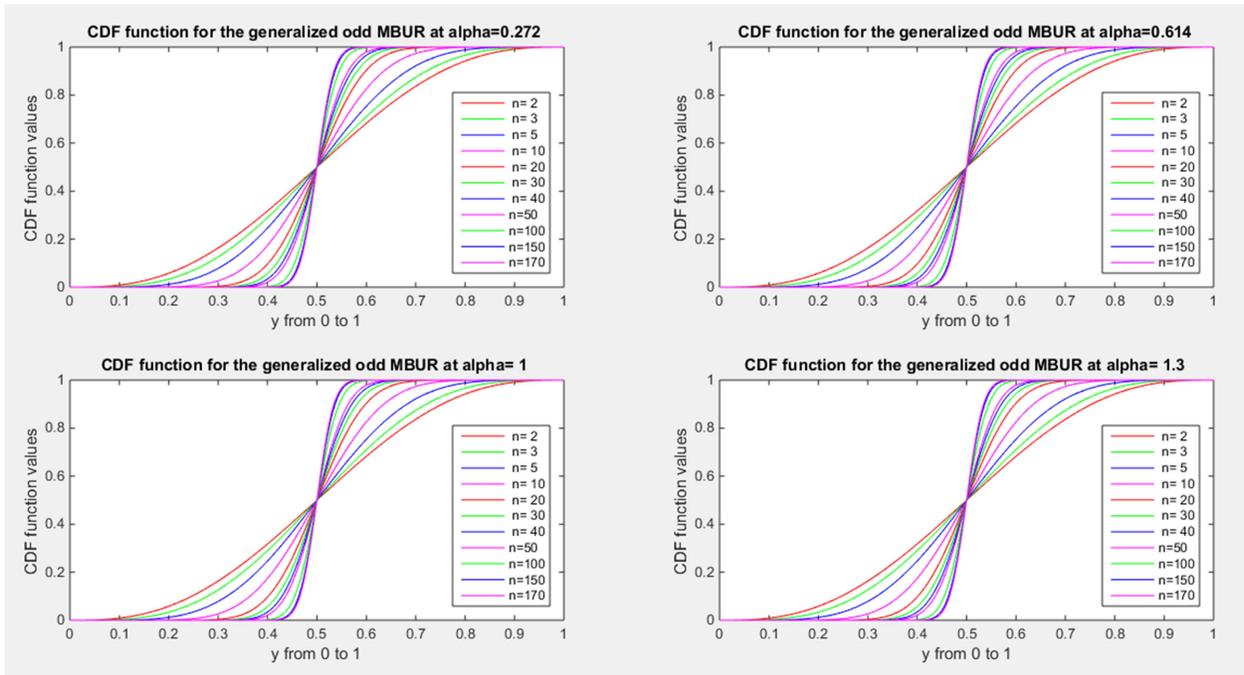

Fig. 20 shows CDF of the first version for different levels of alpha and n

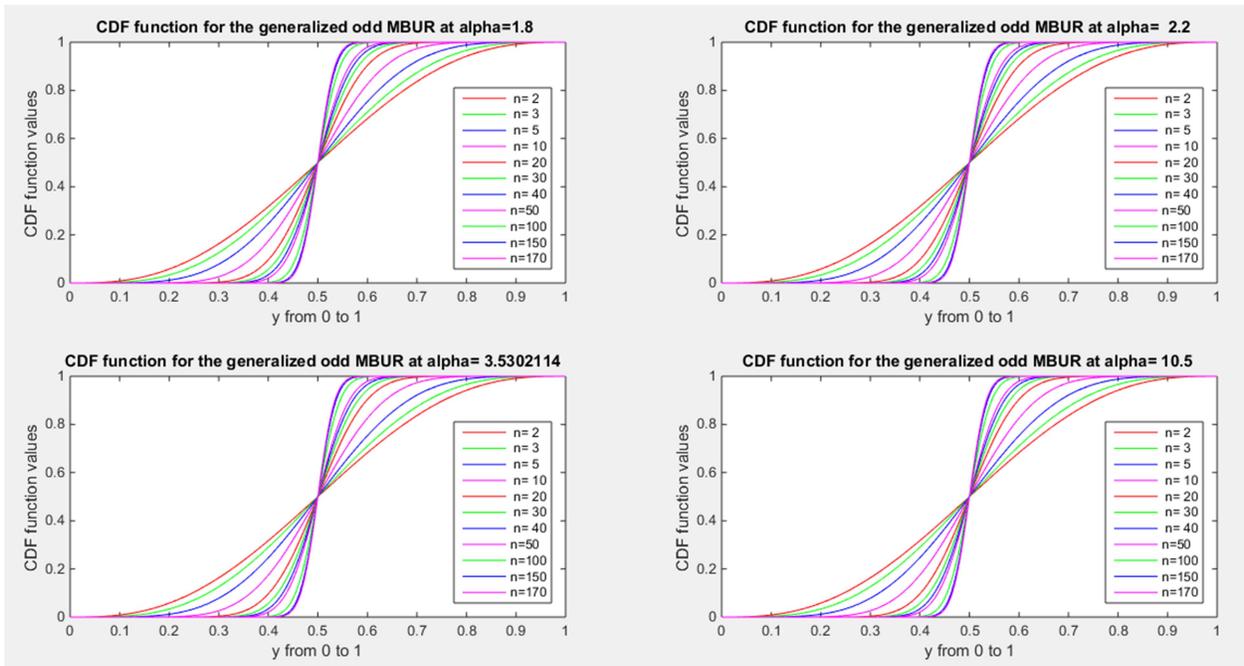

Fig. 21 shows CDF of the first version for different levels of alpha and n

Figures (22-23) illustrate the CDF of the second version for different values of n {2, 3, 5, 10, 20, 30, 40, 50, 100, 150, 170} and different values of alpha {0.272, 0.614, 1, 1.3, 1.8, 2.2, 3.5, 10.5}

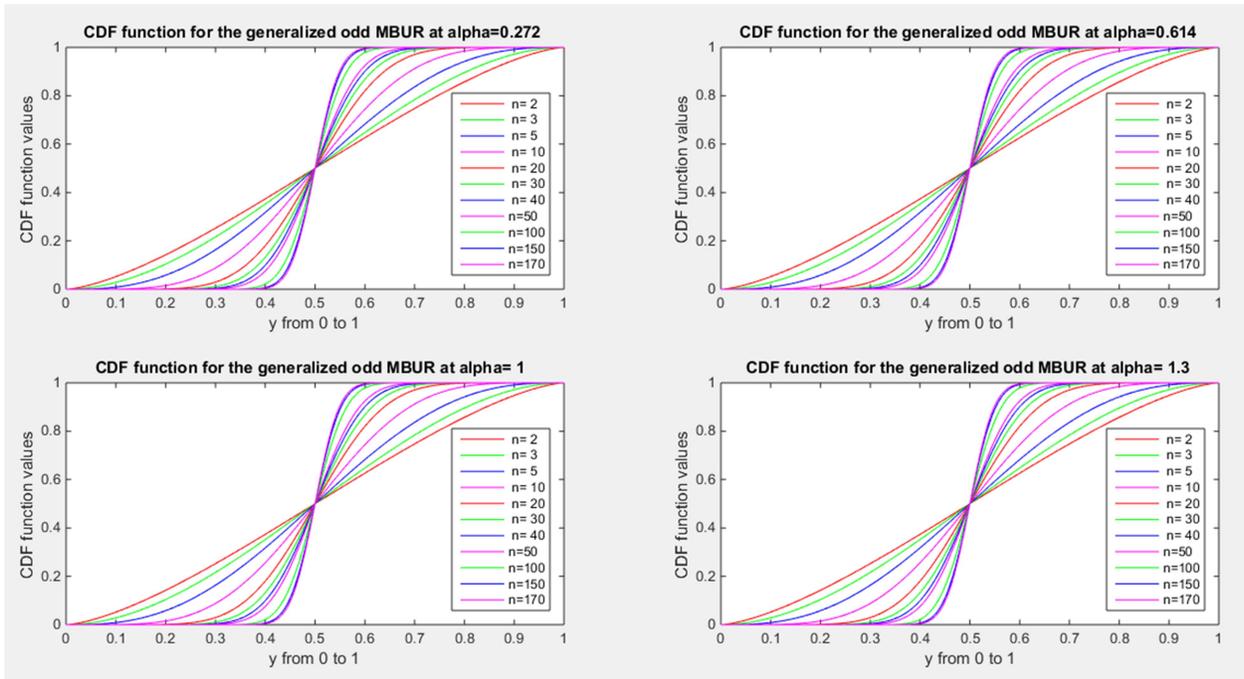

Fig. 22 shows CDF of the second version for different levels of alpha and n

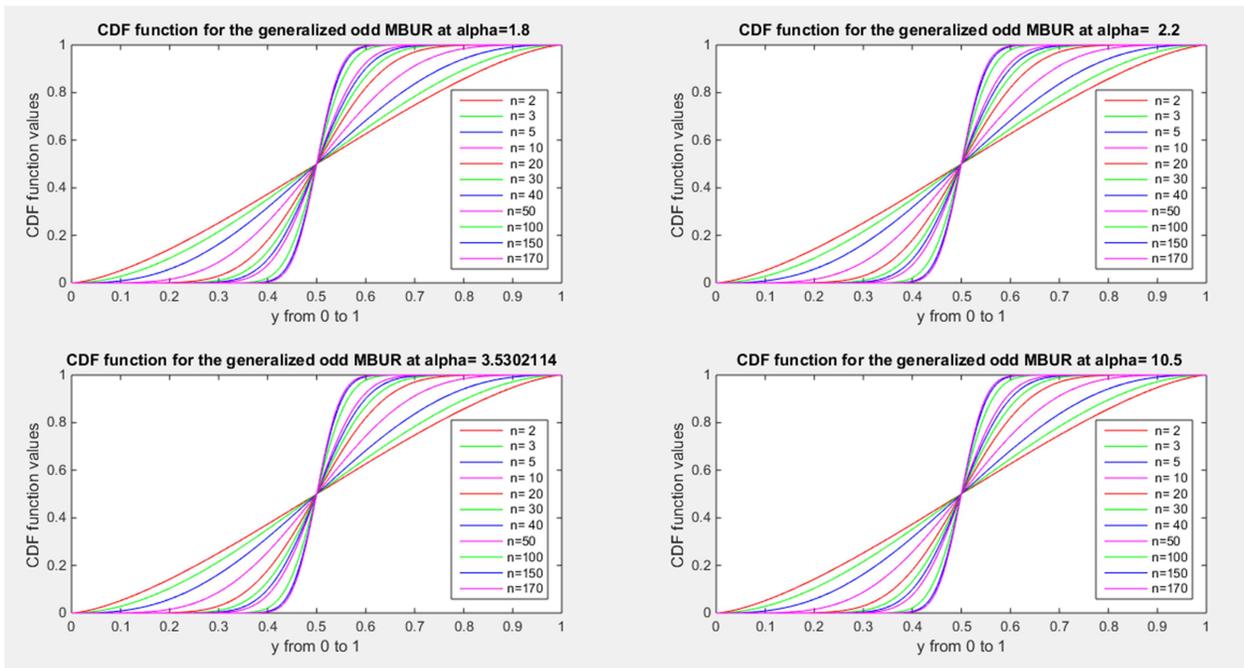

Fig. 23 shows CDF of the second version for different levels of alpha and n

Figures (24-25) illustrate the survival function of the first version for different values of n {2, 3, 5, 10, 20, 30, 40, 50, 100, 150, 170} and different values of alpha {0.272, 0.614, 1, 1.3, 1.8, 2.2, 3.5, 10.5}

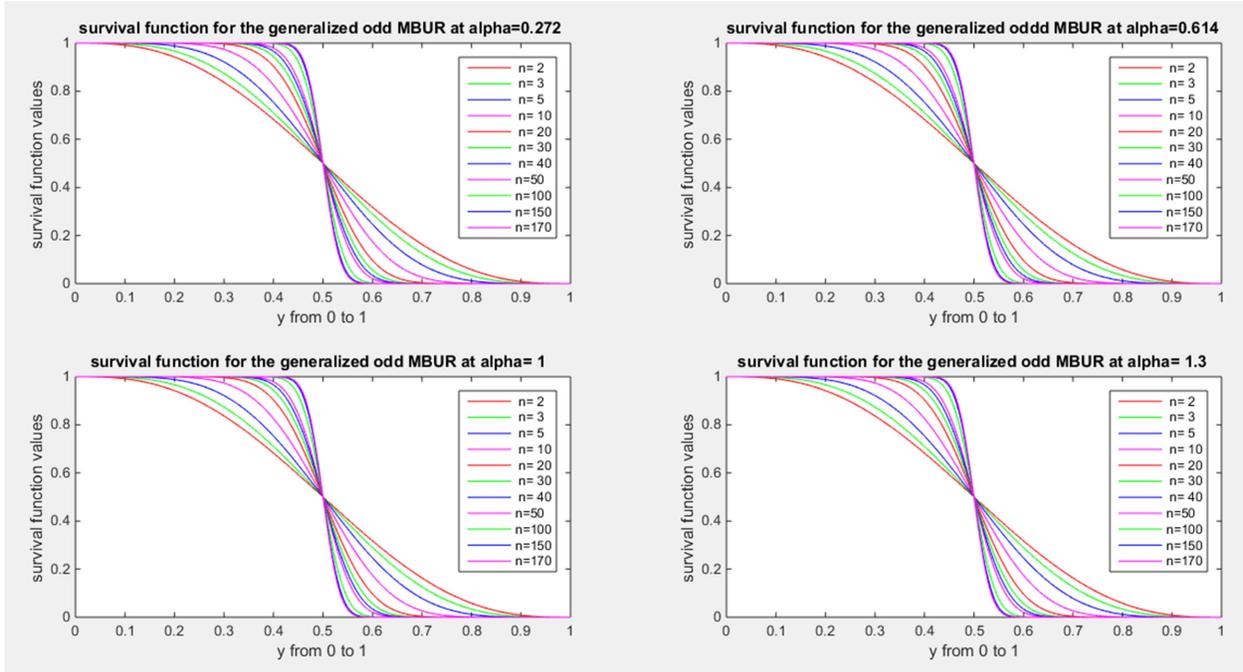

Fig. 24 shows survival function of the first version for different levels of alpha and n

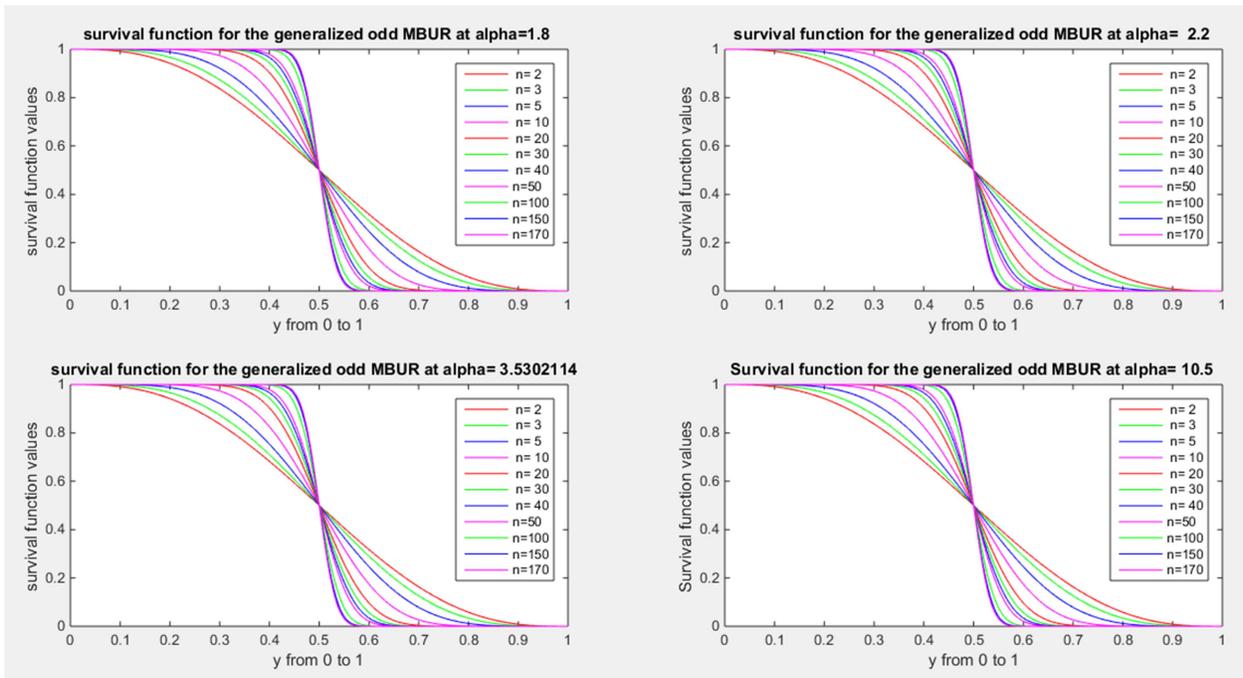

Fig. 25 shows survival function of the first version for different levels of alpha and n

Figures (26-27) illustrate the survival function of the second version for different values of n {2, 3, 5, 10, 20, 30, 40, 50, 100, 150, 170} and different values of alpha {0.272, 0.614, 1, 1.3, 1.8, 2.2, 3.5, 10.5}

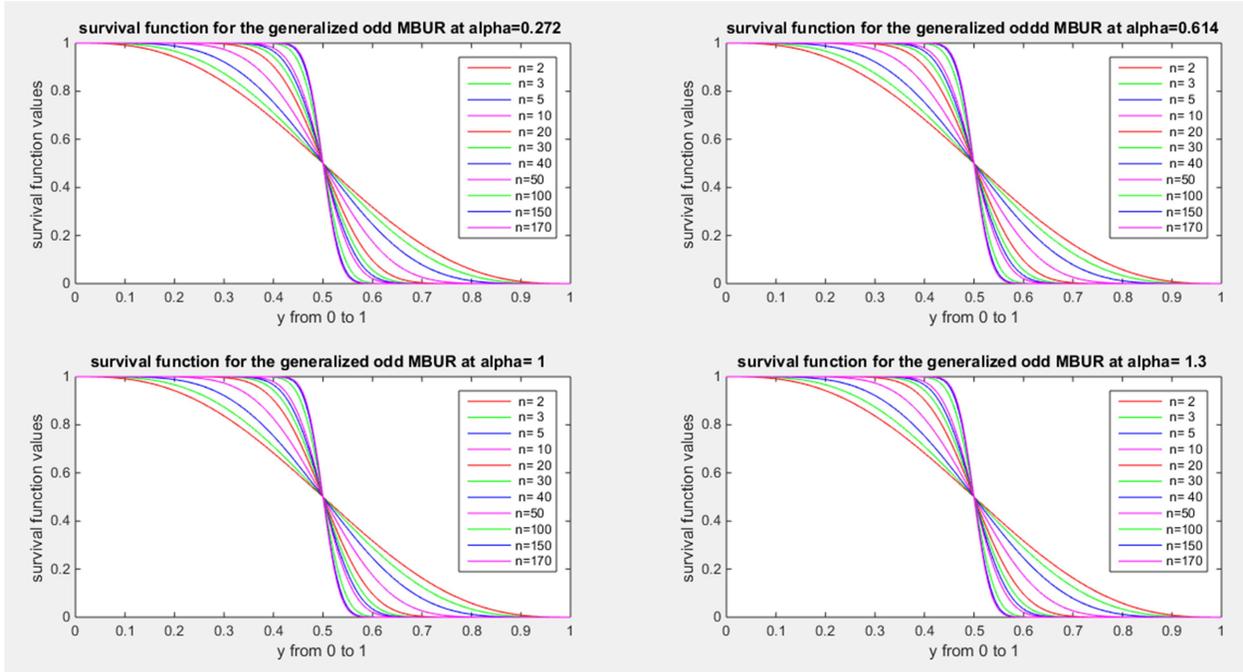

Fig. 26 shows survival function of the second version for different levels of alpha and n

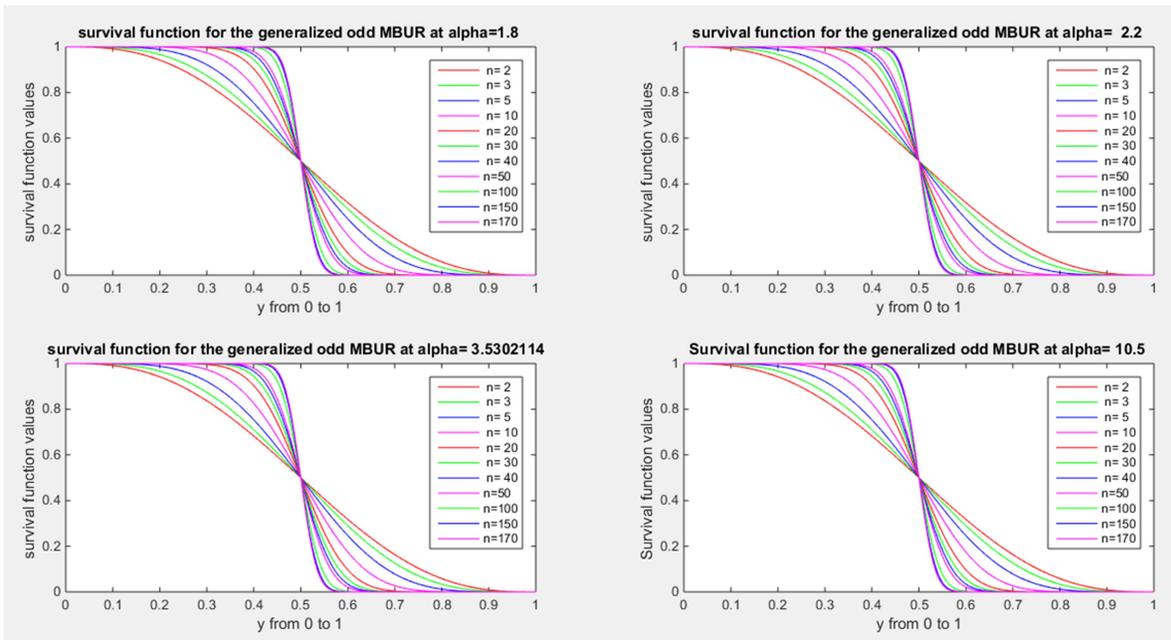

Fig. 27 shows survival function of the second version for different levels of alpha and n

Figures (28-36) illustrate the hazard rate function (hr) of the first version for different values of n {2, 3, 5, 10, 20, 30, 40, 50, 84} and different values of alpha {0.272, 0.614, 1, 1.3, 1.8, 2.2, 3.5, 10.5}

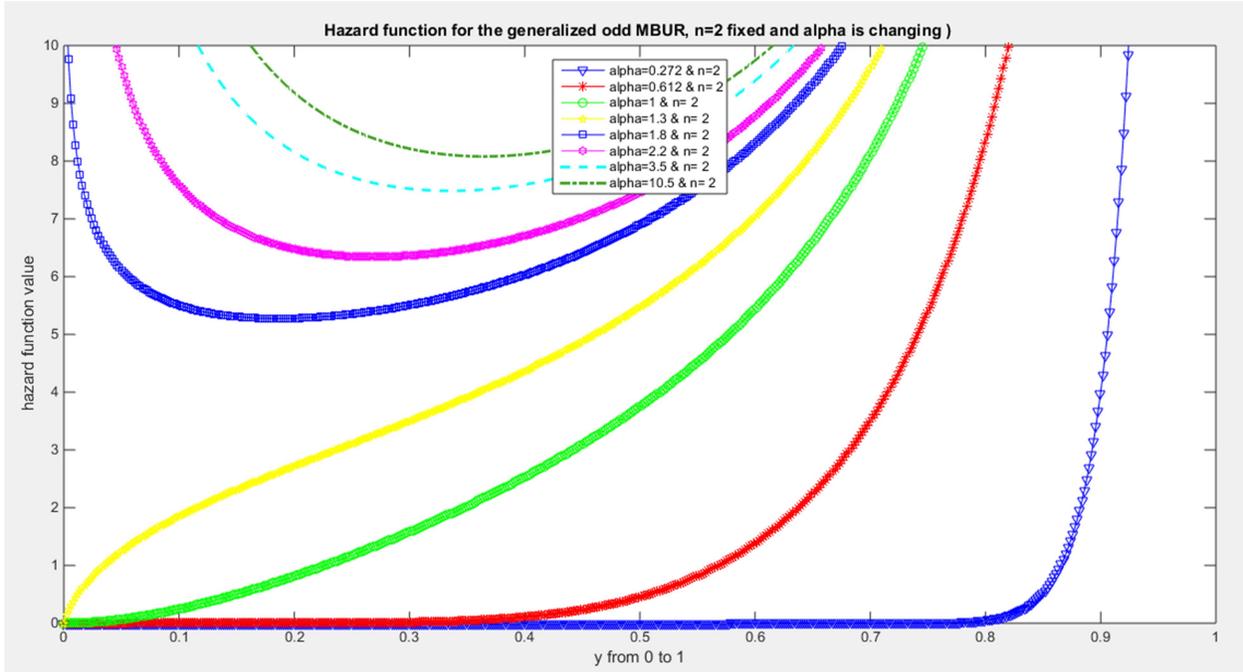

Fig. 28 shows hazard rate function of the first version for different levels of alpha and n=2

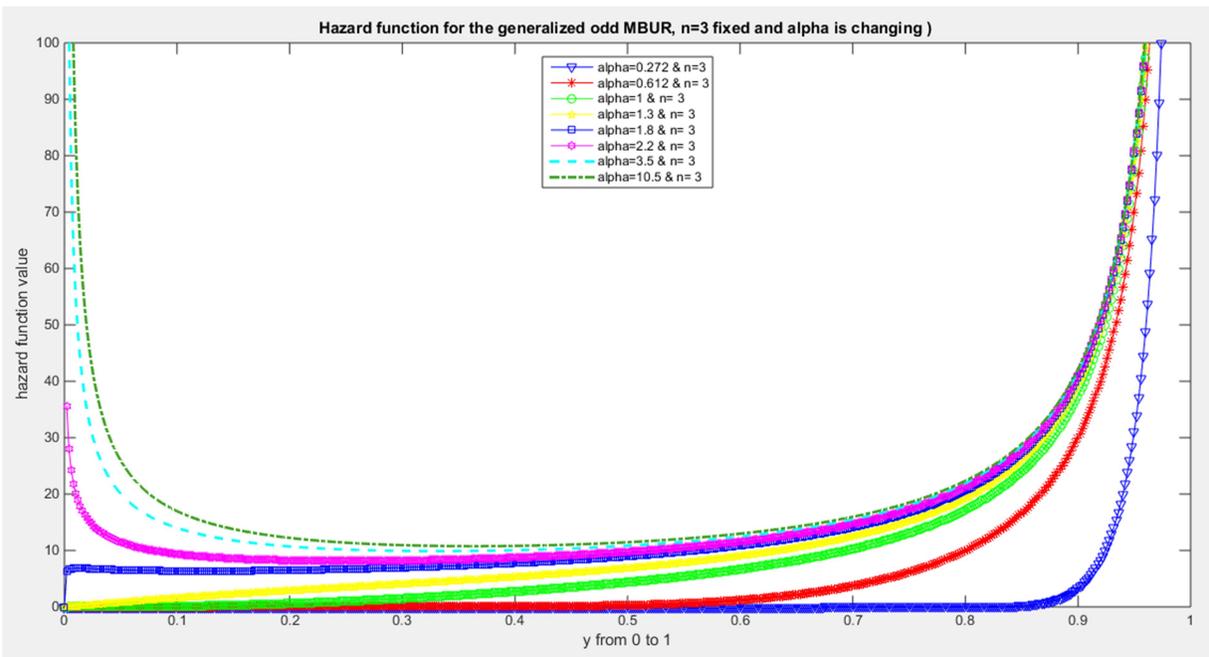

Fig. 29 shows hazard rate function of the first version for different levels of alpha and n=3

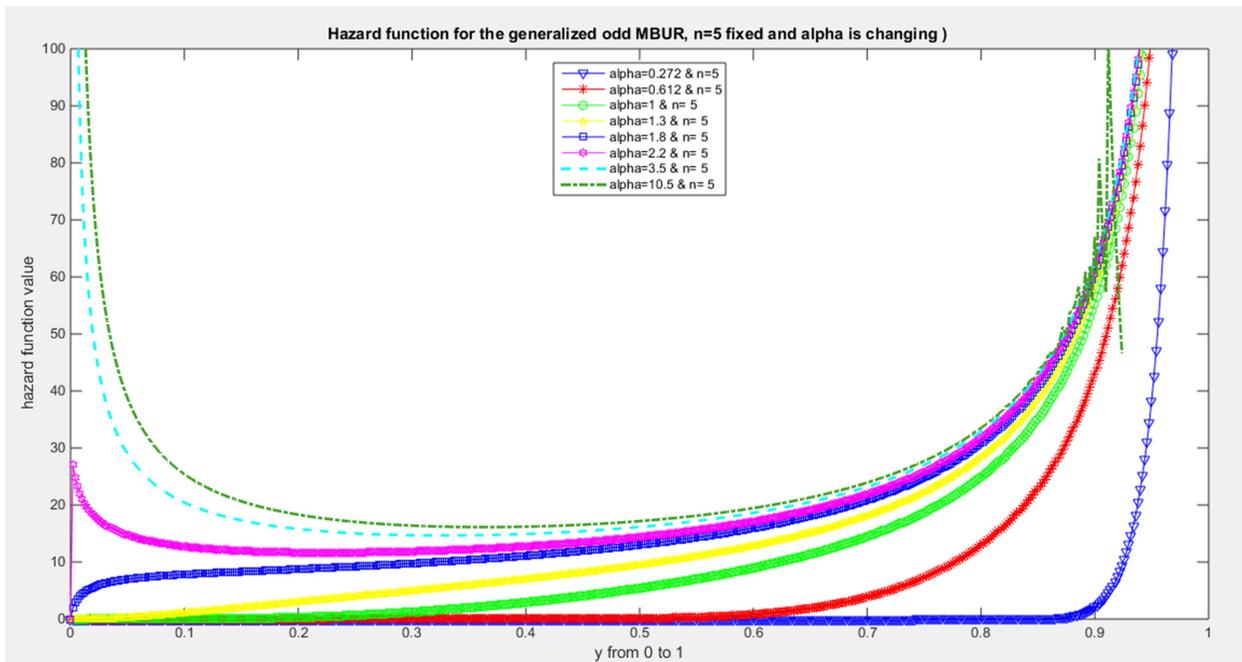

Fig. 30 shows hazard rate function of the first version for different levels of alpha and n=5

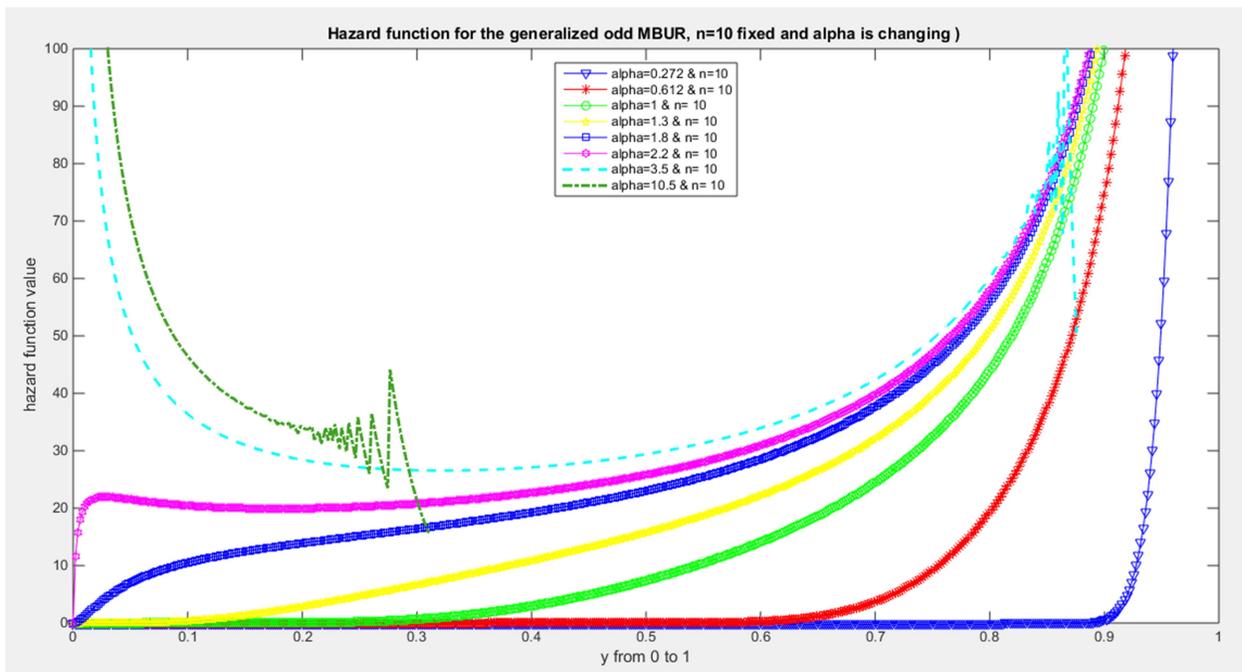

Fig. 31 shows hazard rate function of the first version for different levels of alpha and n=10

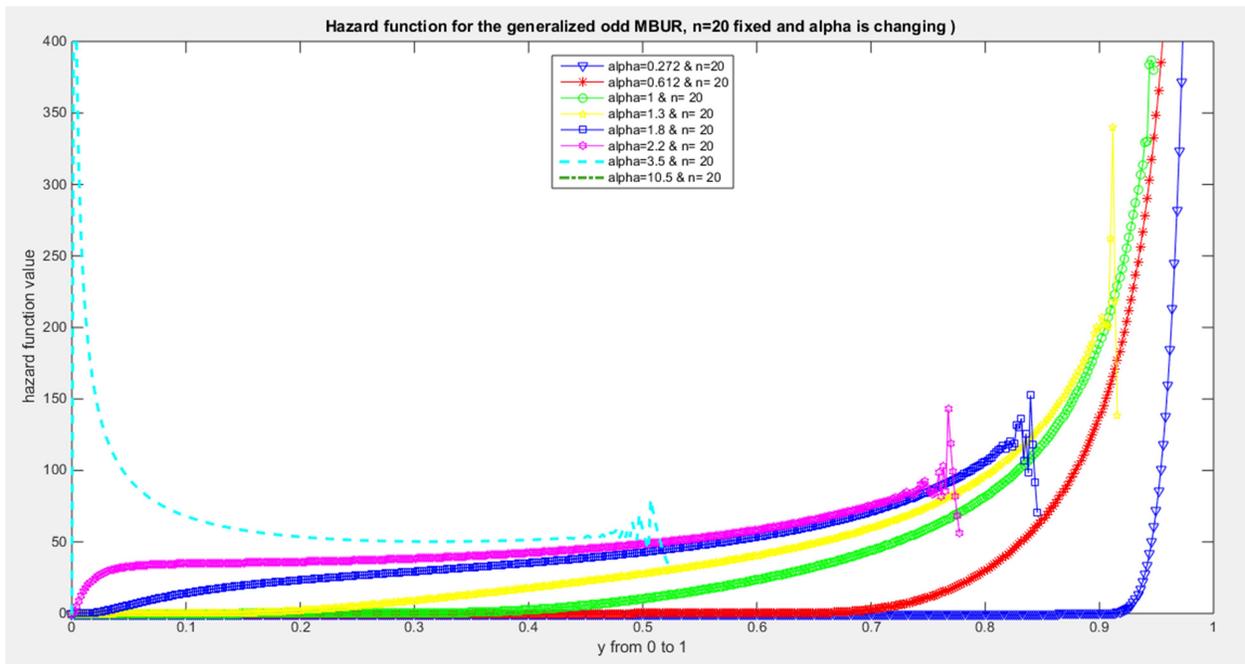

Fig. 32 shows hazard rate function of the first version for different levels of alpha and n=20

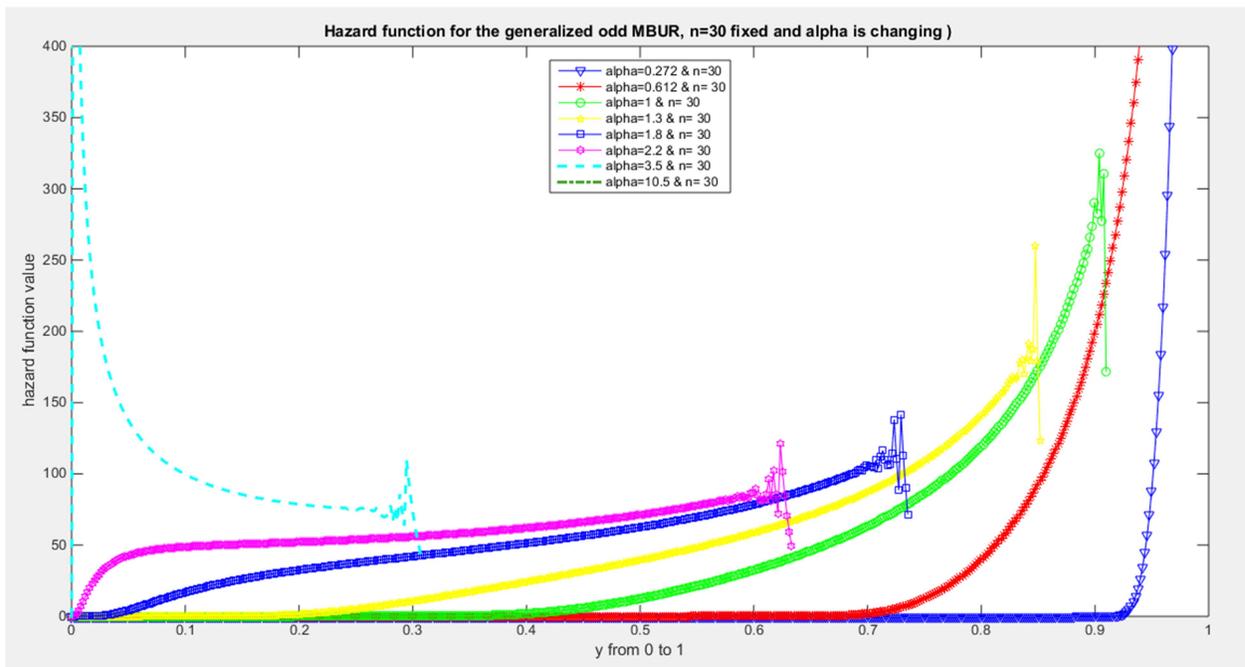

Fig. 33 shows hazard rate function of the first version for different levels of alpha and n=30

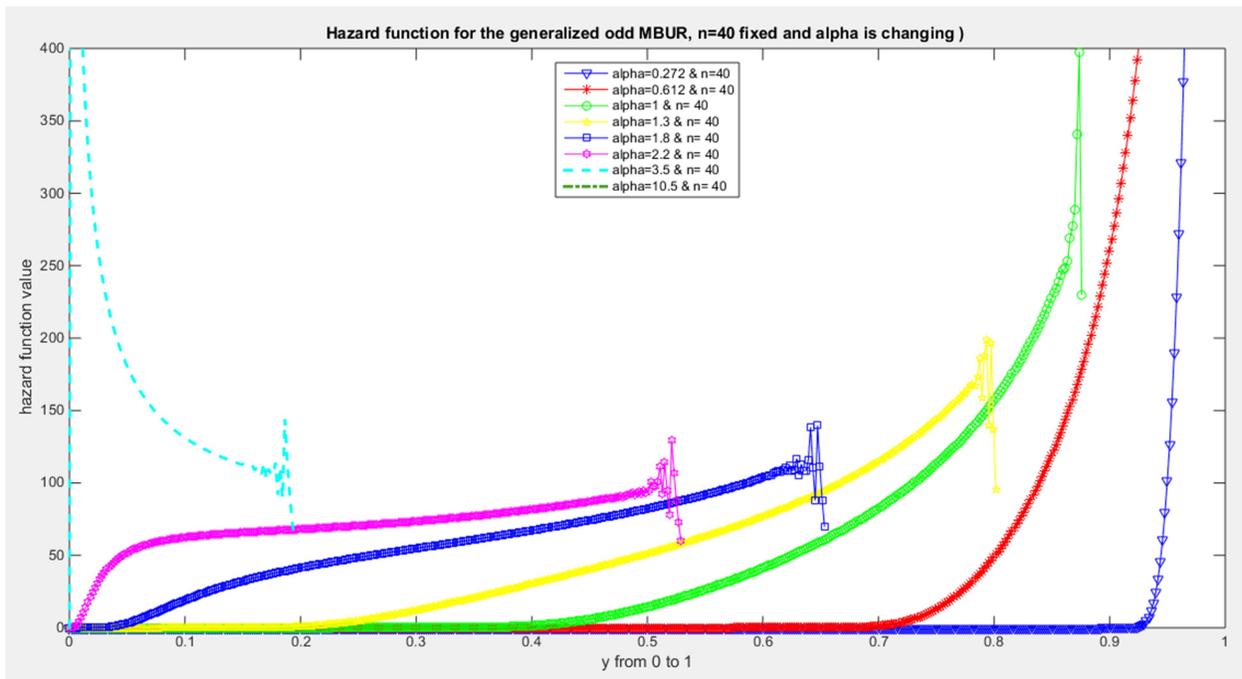

Fig. 34 shows hazard rate function of the first version for different levels of alpha and n=40

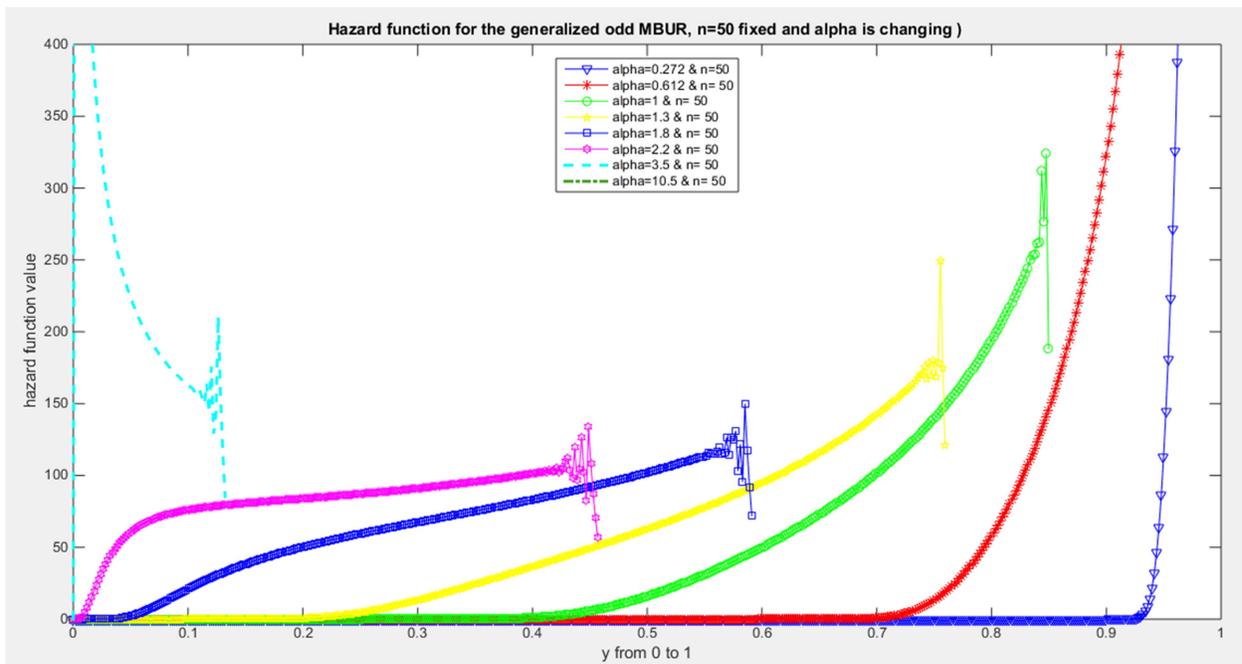

Fig. 35 shows hazard rate function of the first version for different levels of alpha and n=50

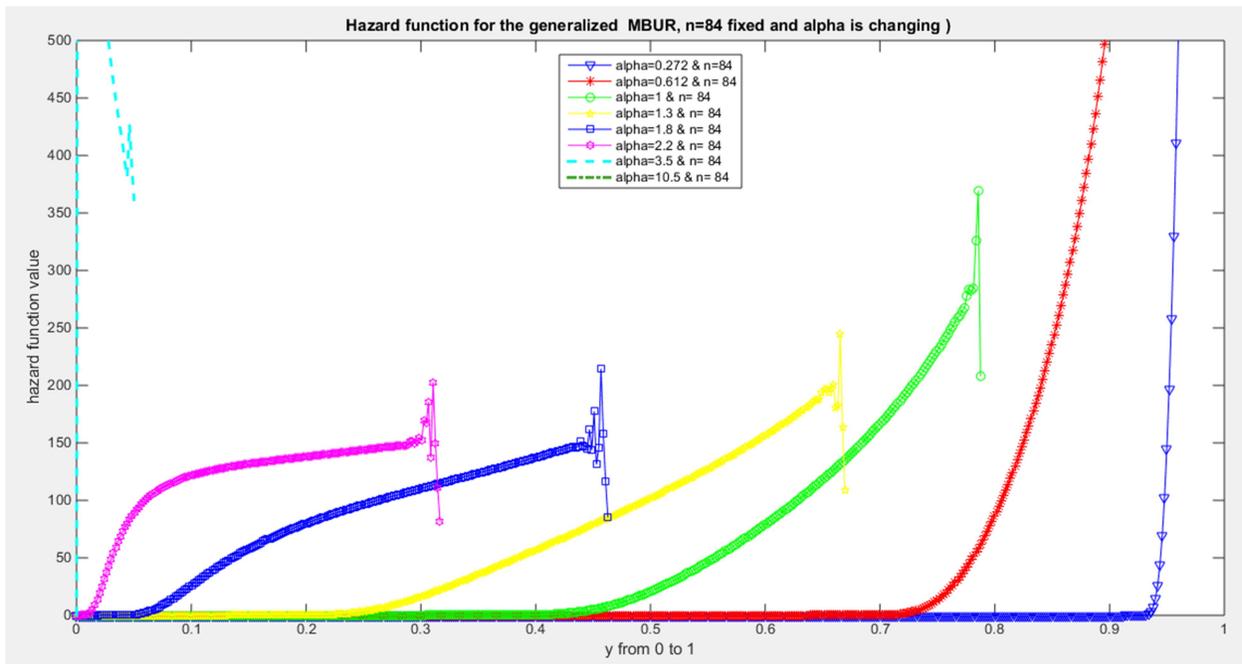

Fig. 36 shows hazard rate function of the first version for different levels of alpha and n=84

Figures (37-47) illustrate the hazard rate function (hr) of the second version for different values of n {2, 3, 5, 10, 20, 30, 40, 50, 100,150,170} and different values of alpha {0.272, 0.614, 1, 1.3, 1.8, 2.2, 3.5, 10.5}

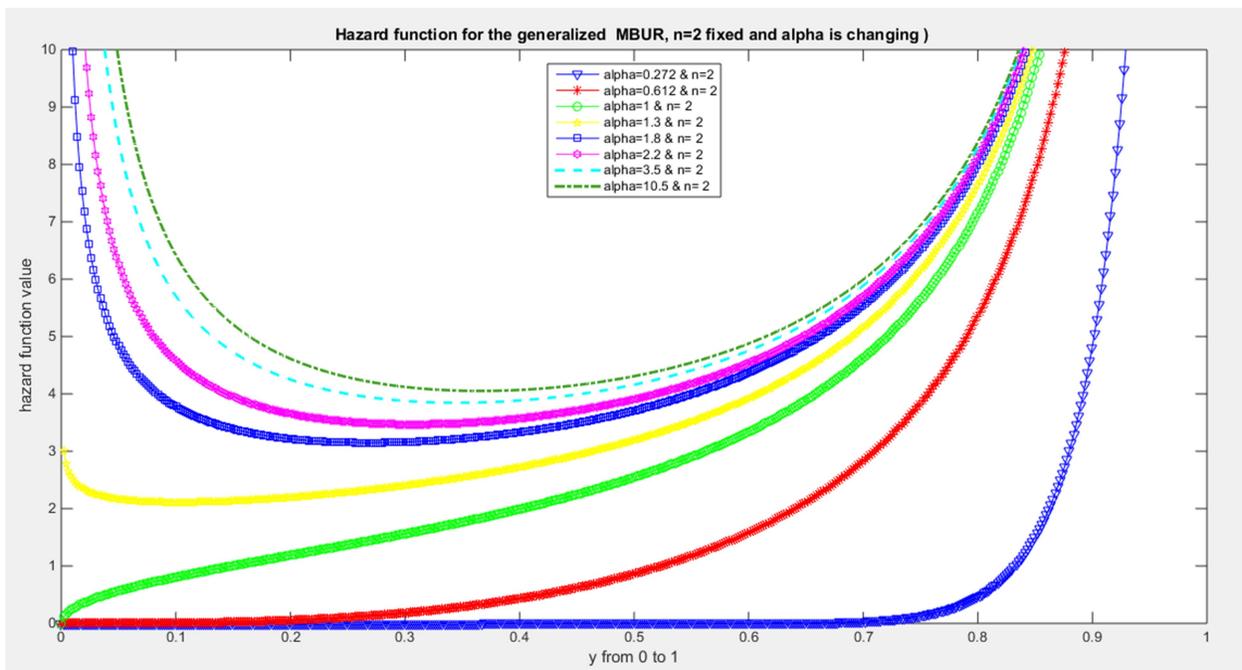

Fig. 37 shows hazard rate function of the second version for different levels of alpha and n=2

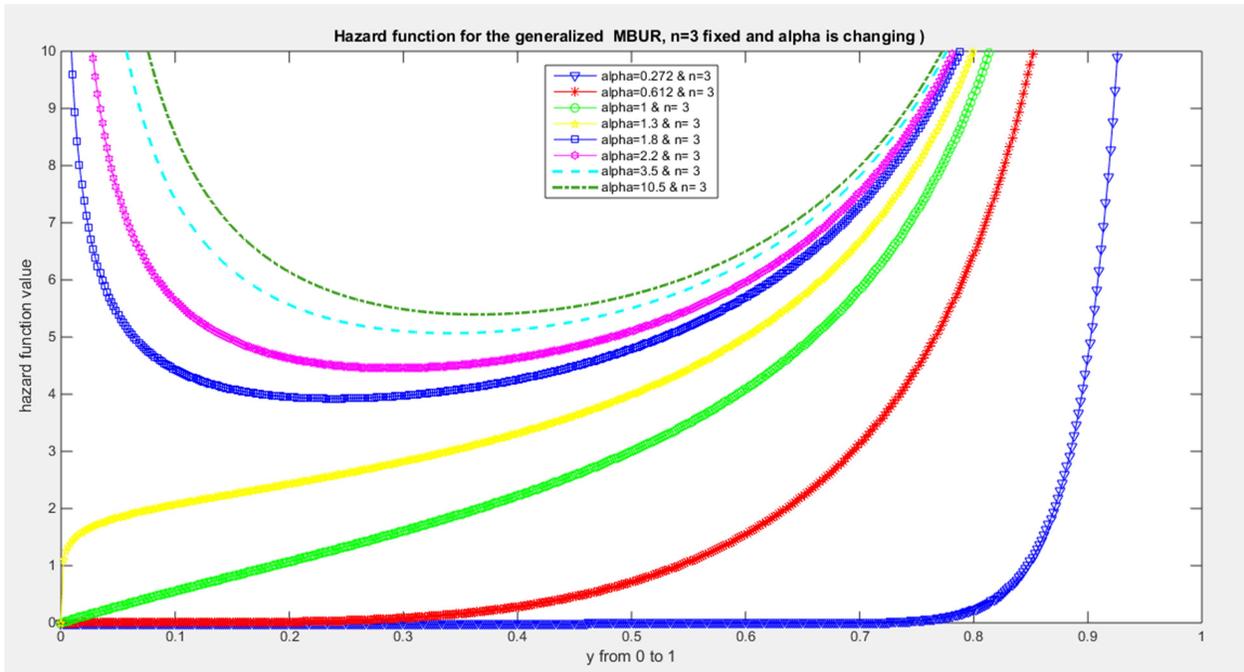

Fig. 38 shows hazard rate function of the second version for different levels of alpha and n=3

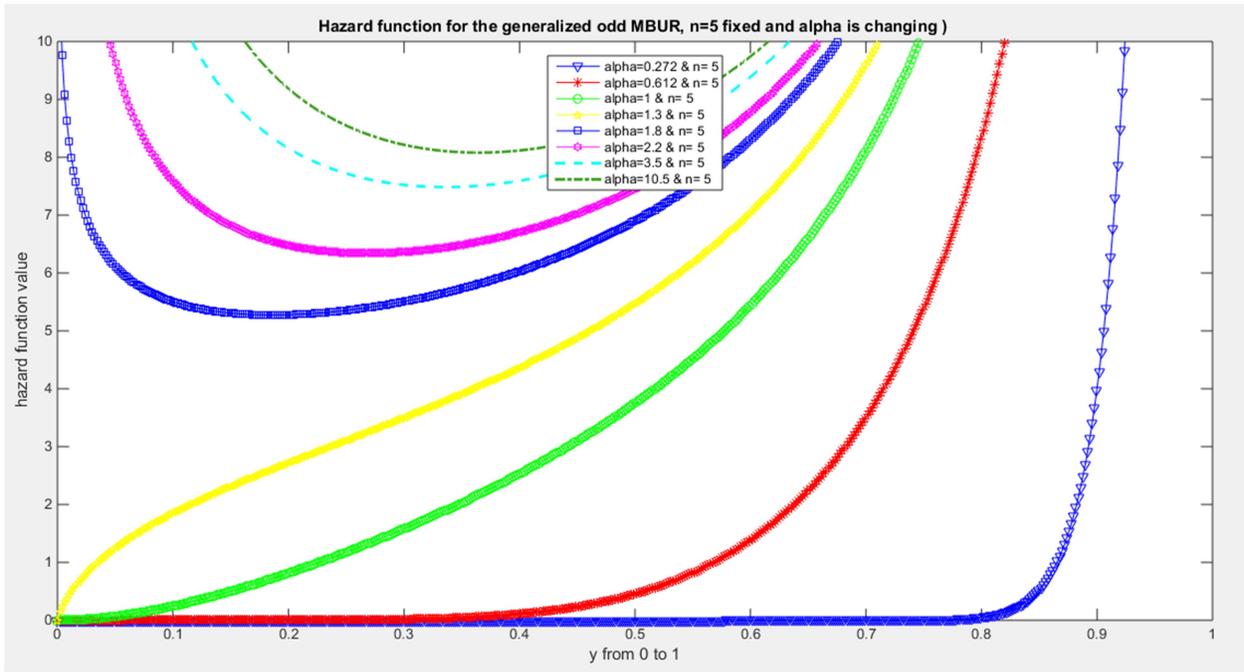

Fig. 39 shows hazard rate function of the second version for different levels of alpha and n=5

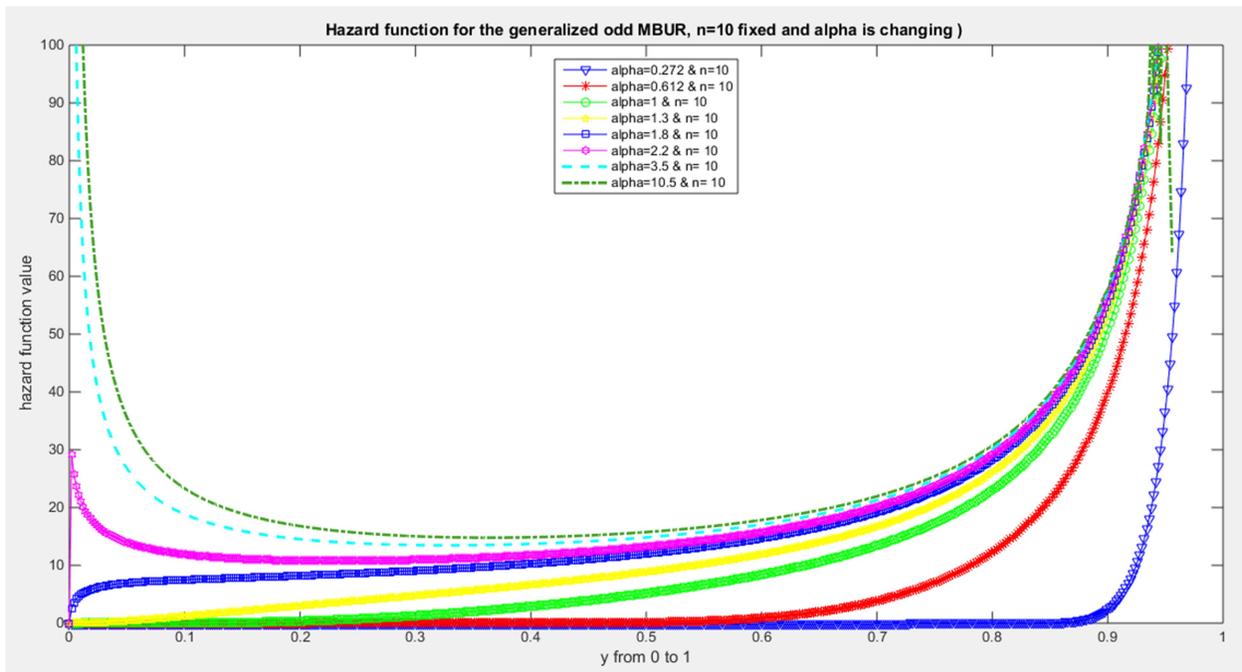

Fig. 40 shows hazard rate function of the second version for different levels of alpha and n=10

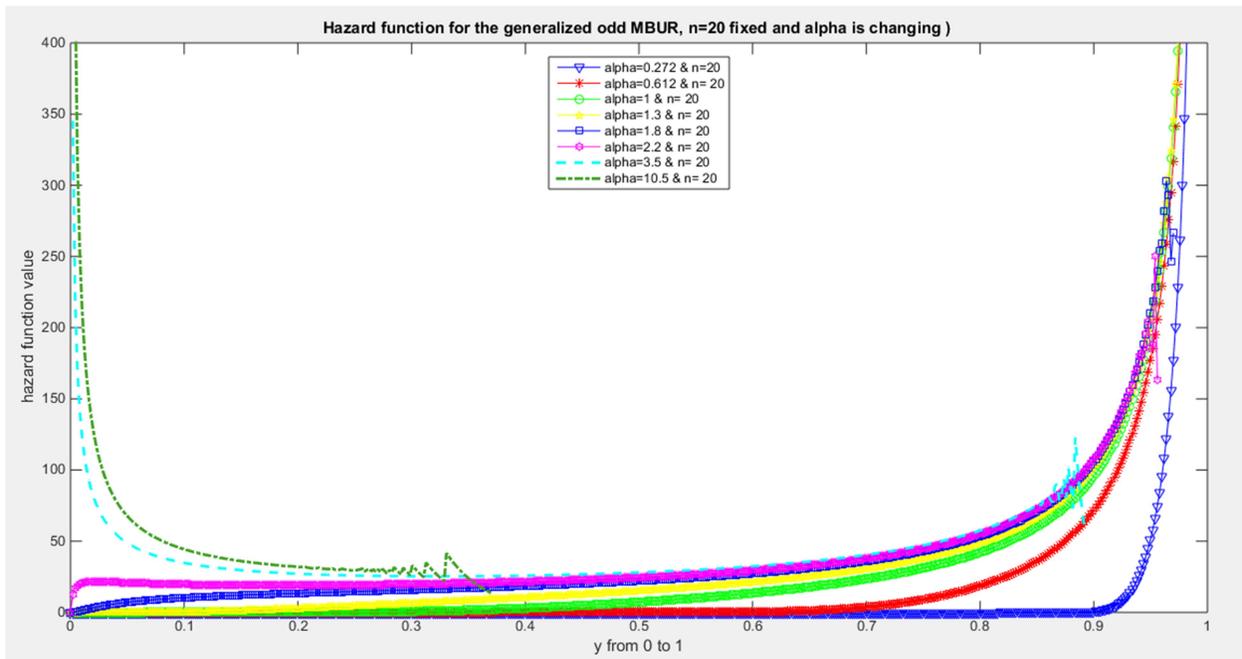

Fig. 41 shows hazard rate function of the second version for different levels of alpha and n=20

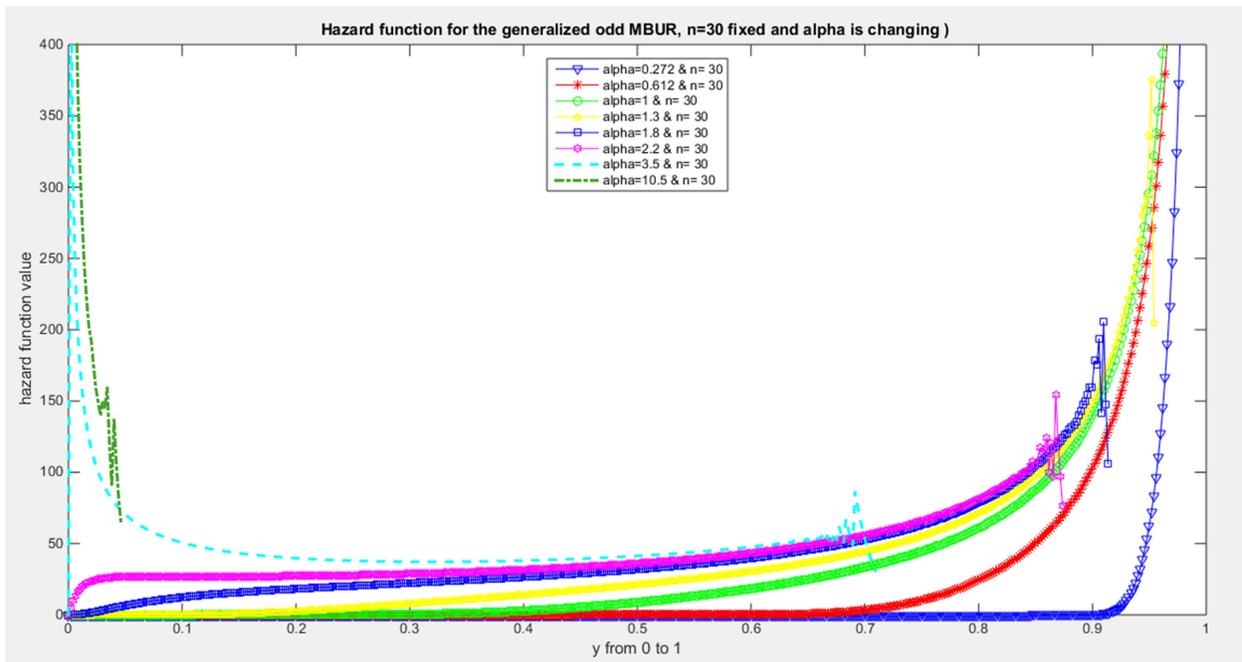

Fig. 42 shows hazard rate function of the second version for different levels of alpha and n=30

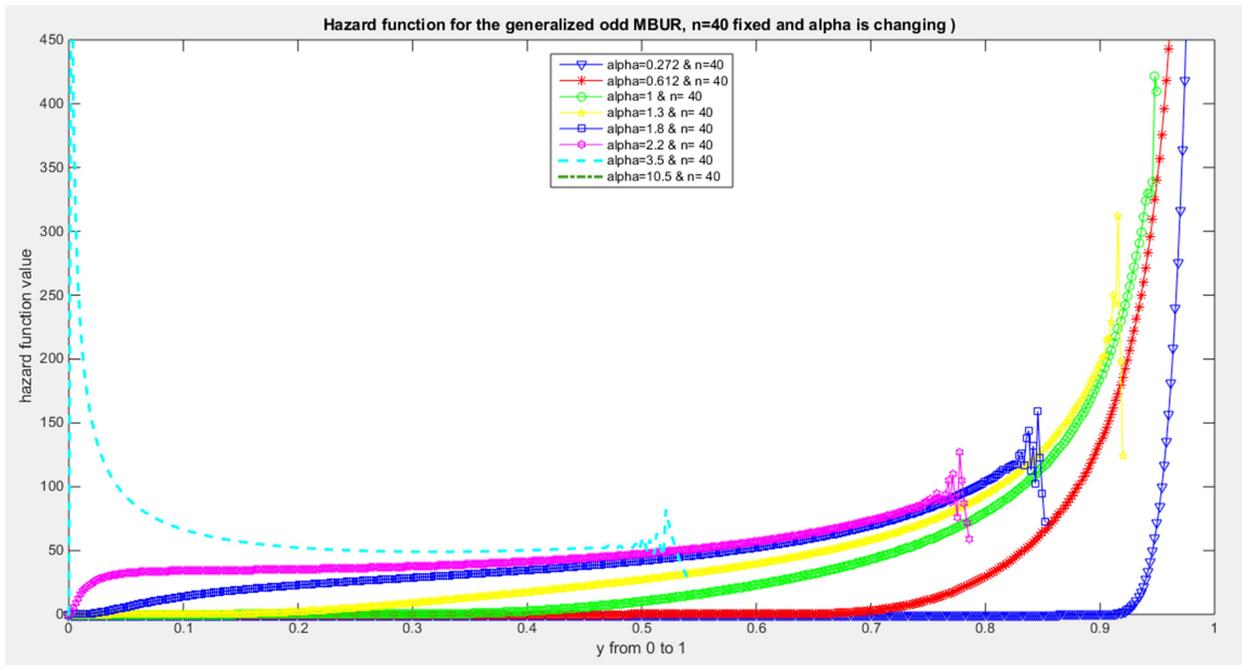

Fig. 43 shows hazard rate function of the second version for different levels of alpha and n=40

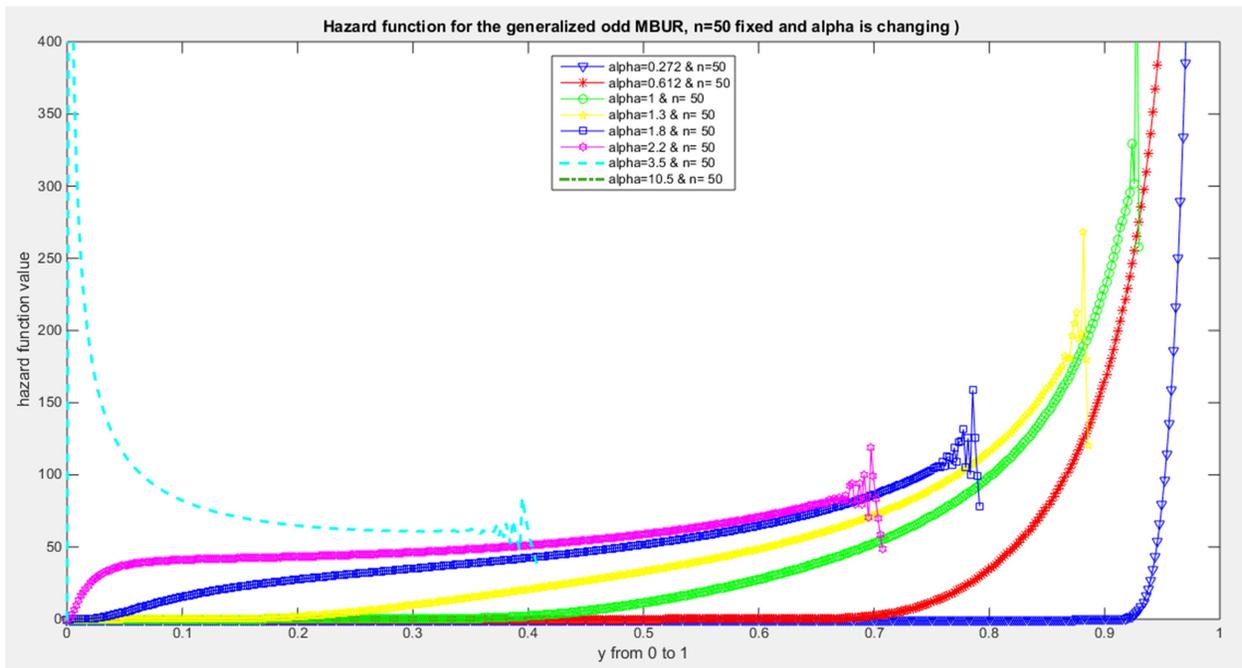

Fig. 44 shows hazard rate function of the second version for different levels of alpha and n=50

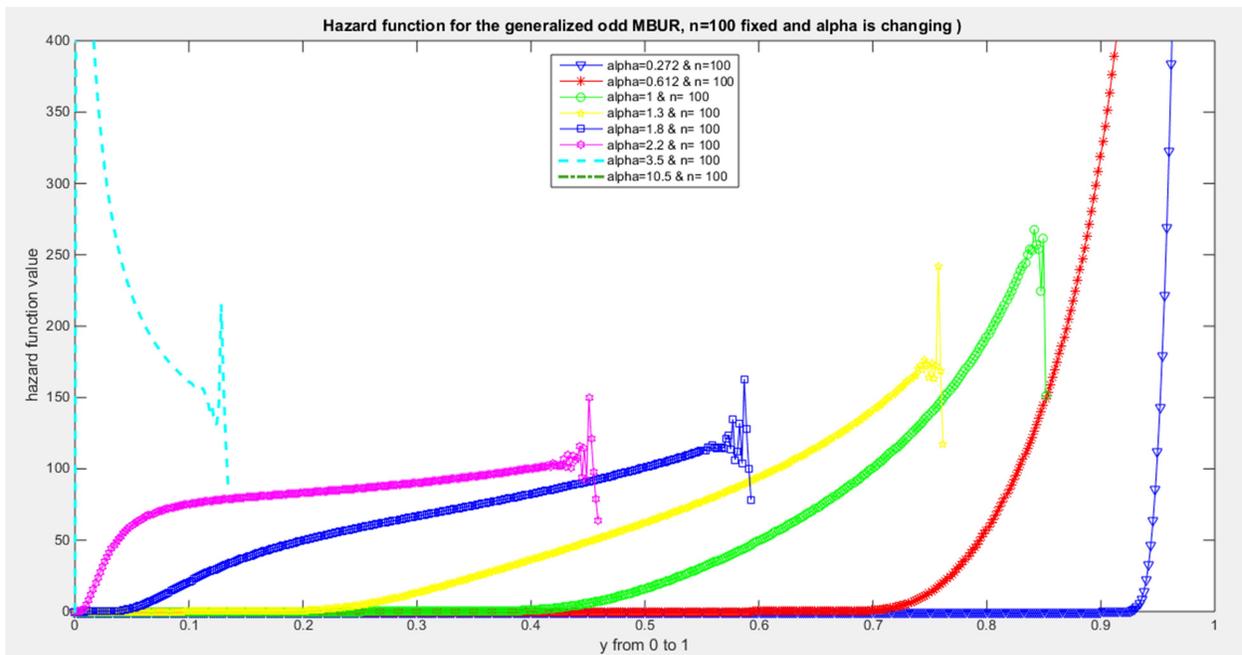

Fig. 45 shows hazard rate function of the second version for different levels of alpha and n=100

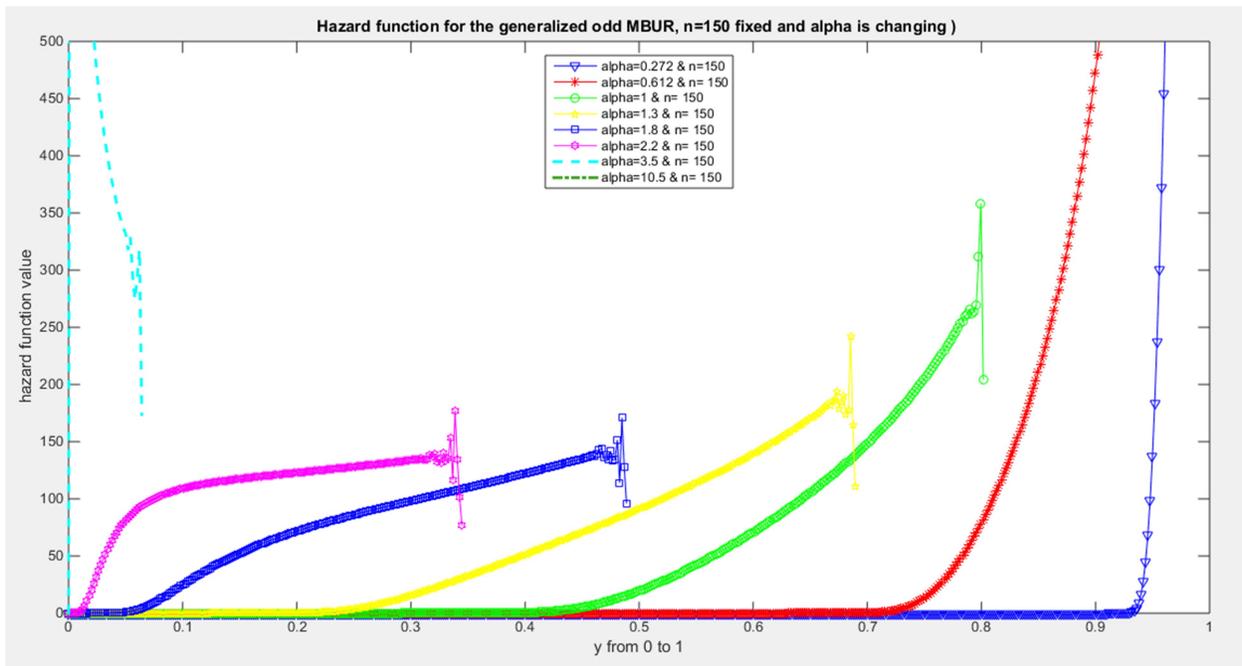

Fig. 46 shows hazard rate function of the second version for different levels of alpha and n=150

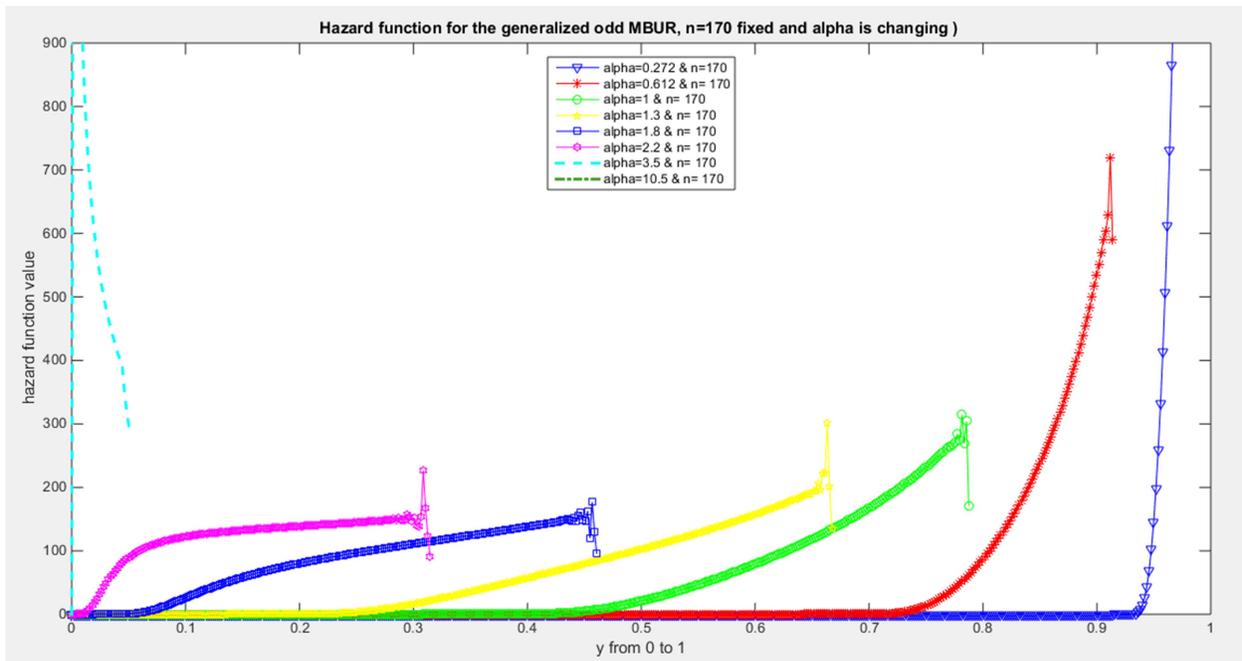

Fig. 47 shows hazard rate function of the second version for different levels of alpha and n=170

The figures of the hazard functions depict the different shapes the hazard function can attain. These shapes ranges from increasing, bath tub and J shaped appearance. The new finding is that when alpha level is large than or equal to one and with increasing n values the hazard rates exhibit oscillating pattern before it starts to approach infinity at the upper end of the unit interval. For example, at the same alpha level 1.8 and with increased values of n, this oscillating pattern is attained at lower values of the random variable . In other words, when n=50, the oscillating pattern is attained at y=0.8, when n=100, this oscillating patter is achieved at y= 0.6, when n=150, the oscillating pattern is captured at y=0.5, when n=170, the oscillating pattern is realized at y values between 0.4 and 0.5. This pattern is almost present at different values of n and alpha. The oscillating pattern in the different graphs abruptly end because the hazard rate values are infinite at these interrupts.

Another new finding is that at higher values of alpha more than two accompanied with higher n values between 20 and 30, the hazard rate function is concave; it starts at zero and then increasing until a specific peak after which it starts to fall down till reaching an oscillating pattern then it starts to approach infinity. This is obvious at alpha=2.2 and when n value is between 20 and 30 and thereafter up to n=150

**Theorem 3:** the rth raw moment of the first version of the distribution is given by

$$E(y^r) = \frac{\Gamma(2n+2)}{\Gamma(n+1)} \frac{\Gamma(n+1+r\alpha^2)}{\Gamma(2n+2+r\alpha^2)}$$

**Proof:** the expectation of the $r^{th}$ moment in equation (23) is obtained with the help of the transformation mentioned in equation (12)

$$E(y^r) = \int_0^1 y^r f_y(y) dy = \frac{\Gamma(2n+2)}{\Gamma(n+1)\Gamma(n+1)} \frac{1}{\alpha^2} \int_0^1 y^r \left[1-y^{\alpha^{-2}}\right]^n [y]^{\frac{n+1}{\alpha^2}-1} dy \quad \dots (23)$$

$$= \frac{\Gamma(2n+2)}{\Gamma(n+1)\Gamma(n+1)} \frac{1}{\alpha^2} \int_0^1 y^r \left[1-y^{\alpha^{-2}}\right]^n [y]^{\frac{n+1}{\alpha^2}-1+r} dy$$

$$= \frac{\Gamma(2n+2)}{\Gamma(n+1)\Gamma(n+1)} \int_0^1 [1-w]^n [w]^{n+r\alpha^2} dw$$

$$= \frac{\Gamma(2n+2)}{\Gamma(n+1)\Gamma(n+1)} \frac{\Gamma(n+1)\Gamma(n+1+r\alpha^2)}{\Gamma(2n+2+r\alpha^2)} = \frac{\Gamma(2n+2)}{\Gamma(n+1)} \frac{\Gamma(n+1+r\alpha^2)}{\Gamma(2n+2+r\alpha^2)}$$

**Theorem 4:** the rth raw moment of the second version of the distribution is given by

$$E(y^r) = \frac{\Gamma(n+1)}{\Gamma\left(\frac{n+1}{2}\right)} \frac{\Gamma\left(\frac{n+1}{2}+r\alpha^2\right)}{\Gamma(n+1+r\alpha^2)}$$

**Proof:** the expectation of the r[th] moment in equation (24) is obtained with the help of the transformation mentioned in equation (12)

$$E(y^r) = \int_0^1 y^r f_y(y) dy = \frac{\Gamma(n+1)}{\Gamma\left(\frac{n}{2}+\frac{1}{2}\right)\Gamma\left(\frac{n}{2}+\frac{1}{2}\right)} \frac{1}{\alpha^2} \int_0^1 y^r \left[1 - y^{\alpha^{-2}}\right]^{\frac{n-1}{2}} [y]^{\frac{n+1}{2\alpha^2}-1} dy \ldots (24)$$

$$= \frac{\Gamma(n+1)}{\Gamma\left(\frac{n}{2}+\frac{1}{2}\right)\Gamma\left(\frac{n}{2}+\frac{1}{2}\right)} \frac{1}{\alpha^2} \int_0^1 \left[1 - y^{\alpha^{-2}}\right]^{\frac{n-1}{2}} [y]^{\frac{n+1}{2\alpha^2}-1+r} dy$$

$$= \frac{\Gamma(n+1)}{\Gamma\left(\frac{n}{2}+\frac{1}{2}\right)\Gamma\left(\frac{n}{2}+\frac{1}{2}\right)} \int_0^1 [1-w]^{\frac{n-1}{2}} [w]^{\frac{n-1}{2}+r\alpha^2} dy$$

$$= \frac{\Gamma(n+1)}{\Gamma\left(\frac{n+1}{2}\right)\Gamma\left(\frac{n+1}{2}\right)} \frac{\Gamma\left(\frac{n+1}{2}\right)\Gamma\left(\frac{n+1}{2}+r\alpha^2\right)}{\Gamma(n+1+r\alpha^2)} = \frac{\Gamma(n+1)}{\Gamma\left(\frac{n+1}{2}\right)} \frac{\Gamma\left(\frac{n}{2}+\frac{1}{2}+r\alpha^2\right)}{\Gamma(n+1+r\alpha^2)}$$

## Section 3: Real Data Analysis

Table 1 shows the flood data. These are 20 observations regarding the maximum flood level of the Susquehanna River at Harrisburg, Pennsylvania. (19).

Table (1): shows Flood Data set

| 0.26 | 0.27 | 0.3  | 0.32 | 0.32 | 0.34 | 0.38 | 0.38 | 0.39 | 0.4  |
|------|------|------|------|------|------|------|------|------|------|
| 0.41 | 0.42 | 0.42 | 0.42 | 045  | 0.48 | 0.49 | 0.61 | 0.65 | 0.74 |

Table 2 shows the descriptive statistics of the data with right skewness and mild positive excess kertosis (leprokurtic).

Table (2): Descriptive statistics of the fourth data set

| min  | mean   | std    | skewness | kurtosis | 25p  | 50p   | 75p   | max  |
|------|--------|--------|----------|----------|------|-------|-------|------|
| 0.26 | 0.4225 | 0.1244 | 1.1625   | 4.2363   | 0.33 | 0.405 | 0.465 | 0.74 |

Table 3 shows the statistical analysis for first version of the Generalized Odd MBUR distribution and its competitors. It outperforms all of them as it has the highest value of log-likelihood and the most negative values for AIC, CAIC, and BIC. The K-S test fails to reject the null hypothesis which supports that the distribution fits the data well with p value 0.3297. Both the AD statistics and the CVM statistics has the lowest values among the other distribution which favors better fit of the data. These are the results for fitting the first version of the distribution.

Table(3): statistical analysis for the flood data set with the first version of the Generalized Odd MBUR distribution:

|  | Beta | | Kumaraswamy | | Generalized odd MBUR | | Topp-Leone | Unit-Lindley |
|---|---|---|---|---|---|---|---|---|
| theta | $\alpha = 6.8318$ | | $\alpha = 3.3777$ | | $n = 8.1044$ | | 2.2413 | 1.6268 |
|  | $\beta = 9.2376$ | | $\beta = 12.0057$ | | $\alpha = 1.1168$ | |  |  |
| Var | 7.22 | 7.2316 | 0.3651 | 2.8825 | 8.0302 | 0.0177 | 0.2512 | 0.0819 |
|  | 7.2316 | 8.0159 | 2.8825 | 29.963 | 0.0177 | 0.0018 |  |  |
| SE | 0.6008 ( alpha) | | 0.1351 (alpha) | | 0.6336 (n) | | 0.1121 | 0.0639 |
|  | 0.6331 ( beta) | | 1.2239 ( beta) | | 0.0095 ( alpha) | |  |  |
| AIC | -24.3671 | | -21.9465 | | -24.4562 | | -12.7627 | -12.3454 |
| CAIC | -23.6613 | | -21.2407 | | -23.7503 | | -12.5405 | -12.1231 |
| BIC | -22.3757 | | -19.9551 | | -22.4647 | | -11.767 | -11.3496 |
| LL | 14.1836 | | 12.9733 | | 14.2281 | | 7.3814 | 7.1727 |
| K-S | 0.2063 | | 0.2175 | | 0.2040 | | 0.3409 | 0.2625 |
| H₀ | Fail to reject | | Fail to reject | | Fail to reject | | Reject | Fail to reject |
| P-value | 0.3174 | | 0.2602 | | 0.3297 | | 0.0141 | 0.0311 |
| AD | 0.7302 | | 0.9365 | | 0.7153 | | 2.9131 | 2.3153 |
| CVM | 0.1242 | | 0.1653 | | 0.1205 | | 0.5857 | 0.4428 |

The P-values for the estimators of alpha and beta parameters of the Beta distribution and Kumaraswamy distributions are significant ($p < 0.001$).

P-values for the estimators of alpha of the MBUR distribution is significant ($p < 0.001$).

P-values for the estimators of theta of the Unit Lindley distribution is significant ($p < 0.001$).

The generalization of the MBUR improves the statistical analysis of the previously analyzed data with the MBUR. It leverages the indices and this generalization makes the MBUR outperforms the beta and the Kumaraswamy distribution. Figure 48 shows the histogram and the fitted Generalized Odd MBUR which shows marked enhancement for fitting the data than the MBUR shown in figure 49. This is also true if comparing the theoretical CDF of the Generalized Odd MBUR in figure 50 than the one in figure 51.

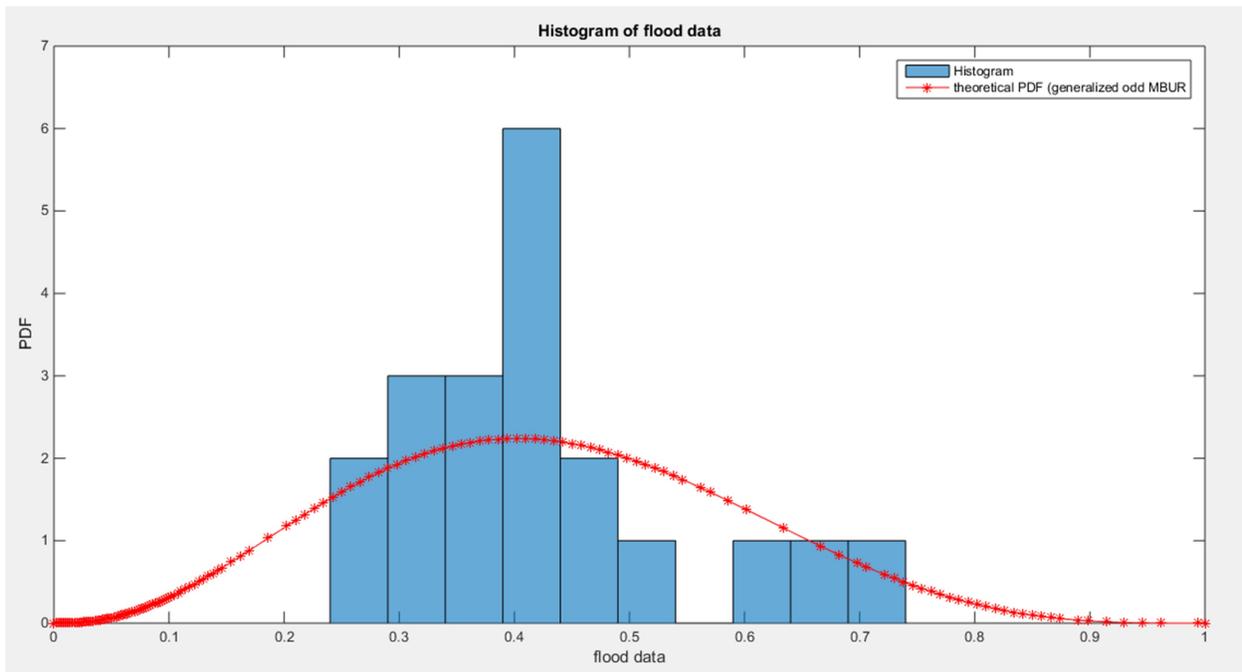

Fig 48 shows the histogram of the flood data and the fitted Generalized Odd MBUR distribution ( first version ) .

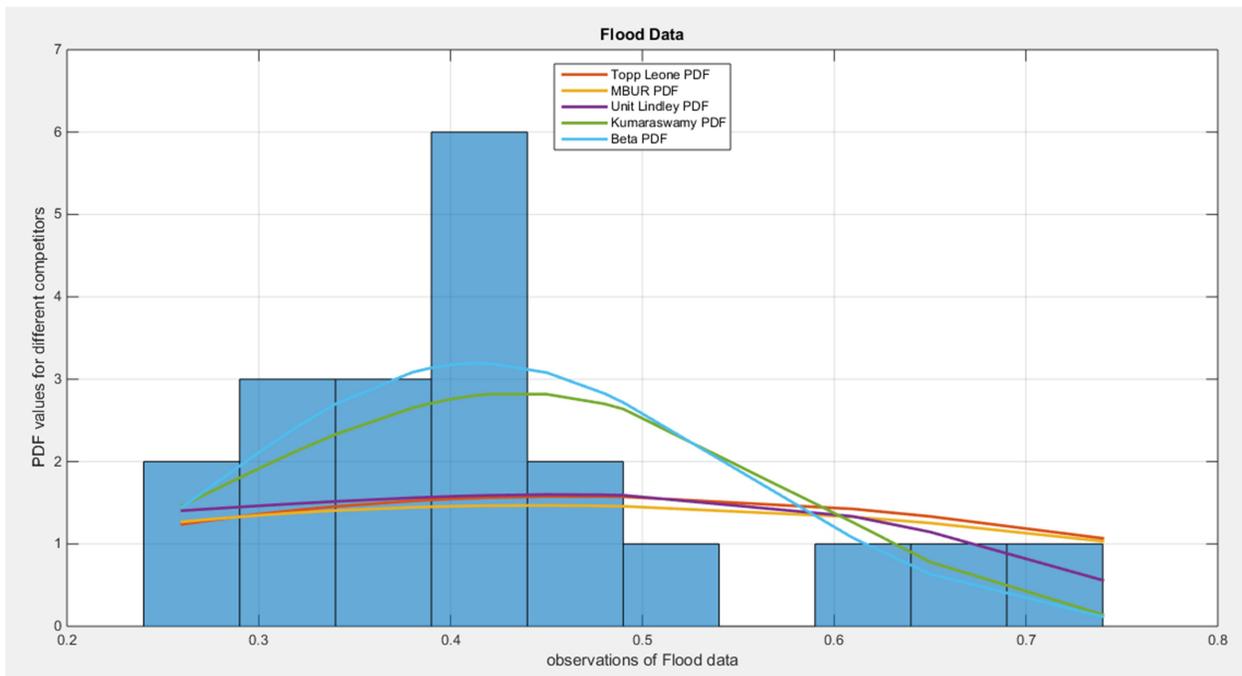

Fig 49 shows the histogram of the flood data and the fitted MBUR distribution and other competitors.

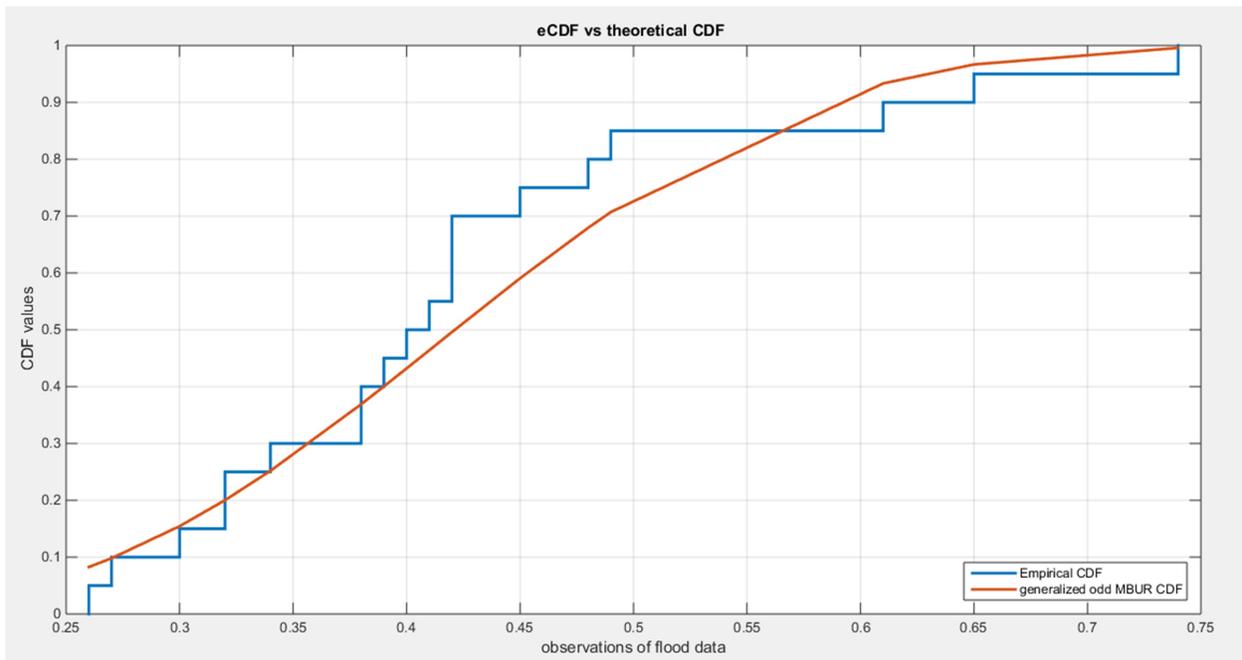

Fig 50 shows the empirical and the theoretical CDF after fitting the generalized Odd MBUR distribution (first version).

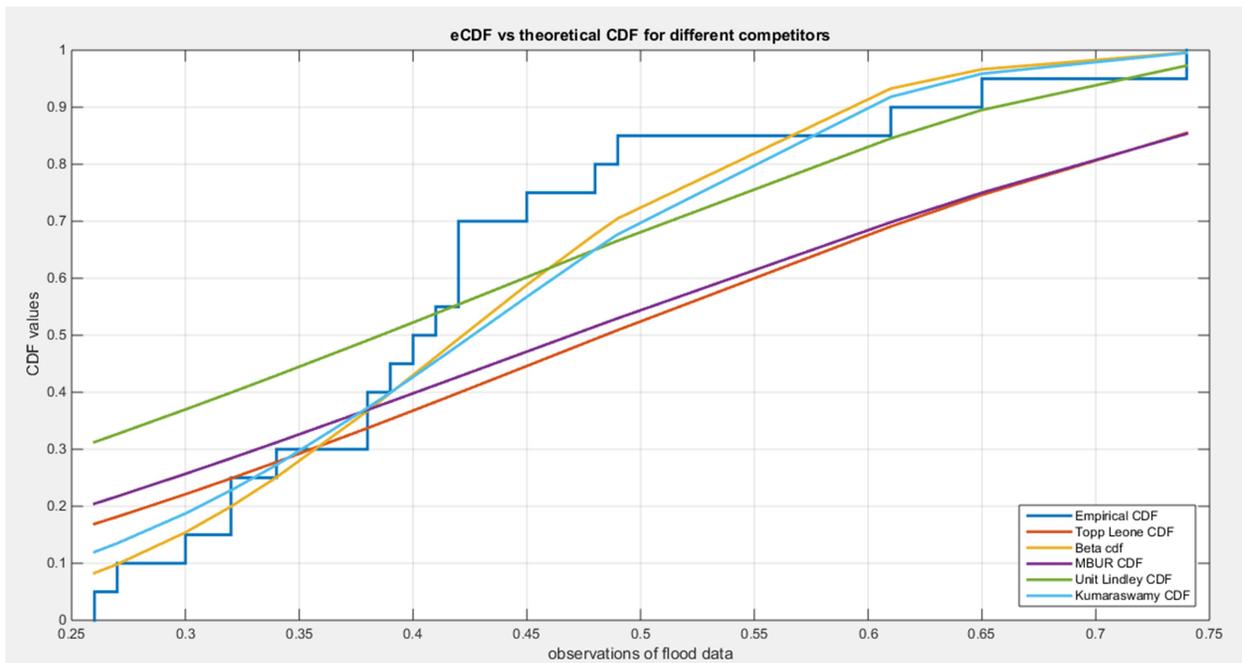

Fig. 51 shows the eCDF vs. theoretical CDF of the distributions for the flood data set. The purple curve is the MBUR. The generalization of MBUR markedly improves fitting the data as shown in figure 50.

Table 4 shows the statistical analysis for second version of the Generalized Odd MBUR distribution and its competitors. The same results are obtained apart from the estimated n parameter value and its associated variance and standard error. Otherwise, it outperforms the other distributions for the same reasons as the first version of it. The value of the estimated n parameter is larger (17.2087) compared with (8.0302) while the value of the estimated alpha is the same(1.1168).

Table (4): statistical analysis for the flood data with the second version of the Generalized Odd MBUR

|  | Beta | | Kumaraswamy | | Generalized odd MBUR | | Topp-Leone | Unit-Lindley |
|---|---|---|---|---|---|---|---|---|
| theta | $\alpha = 6.8318$ | | $\alpha = 3.3777$ | | $n = 17.2087$ | | 2.2413 | 1.6268 |
|  | $\beta = 9.2376$ | | $\beta = 12.0057$ | | $\alpha = 1.1168$ | |  |  |
| Var | 7.22 | 7.2316 | 0.3651 | 2.8825 | 32.1208 | 0.0354 | 0.2512 | 0.0819 |
|  | 7.2316 | 8.0159 | 2.8825 | 29.963 | 0.0354 | 0.0018 |  |  |
| SE | 0.6008 ( alpha) | | 0.1351 (alpha) | | 1.2673 (n) | | 0.1121 | 0.0639 |
|  | 0.6331 ( beta) | | 1.2239 ( beta) | | 0.0095 ( alpha) | |  |  |
| AIC | -24.3671 | | -21.9465 | | -24.4562 | | -12.7627 | -12.3454 |
| CAIC | -23.6613 | | -21.2407 | | -23.7503 | | -12.5405 | -12.1231 |
| BIC | -22.3757 | | -19.9551 | | -22.4647 | | -11.767 | -11.3496 |
| LL | 14.1836 | | 12.9733 | | 14.2281 | | 7.3814 | 7.1727 |
| K-S | 0.2063 | | 0.2175 | | 0.2040 | | 0.3409 | 0.2625 |
| H₀ | Fail to reject | | Fail to reject | | Fail to reject | | Reject | Fail to reject |
| P-value | 0.3174 | | 0.2602 | | 0.3297 | | 0.0141 | 0.0311 |
| AD | 0.7302 | | 0.9365 | | 0.7153 | | 2.9131 | 2.3153 |
| CVM | 0.1242 | | 0.1653 | | 0.1205 | | 0.5857 | 0.4428 |

The P-values for the estimators of alpha and beta parameters of the Beta distribution and Kumaraswamy distributions are significant ($p < 0.001$).

P-values for the estimators of alpha of the MBUR distribution is significant ($p < 0.001$).

P-values for the estimators of theta of the Unit Lindley distribution is significant ($p < 0.001$).

Figure 52 shows the fitted PDF for the second version with slight difference than figure 48. The peak is larger than in figure 48. Figure 53 shows the CDF for this version.

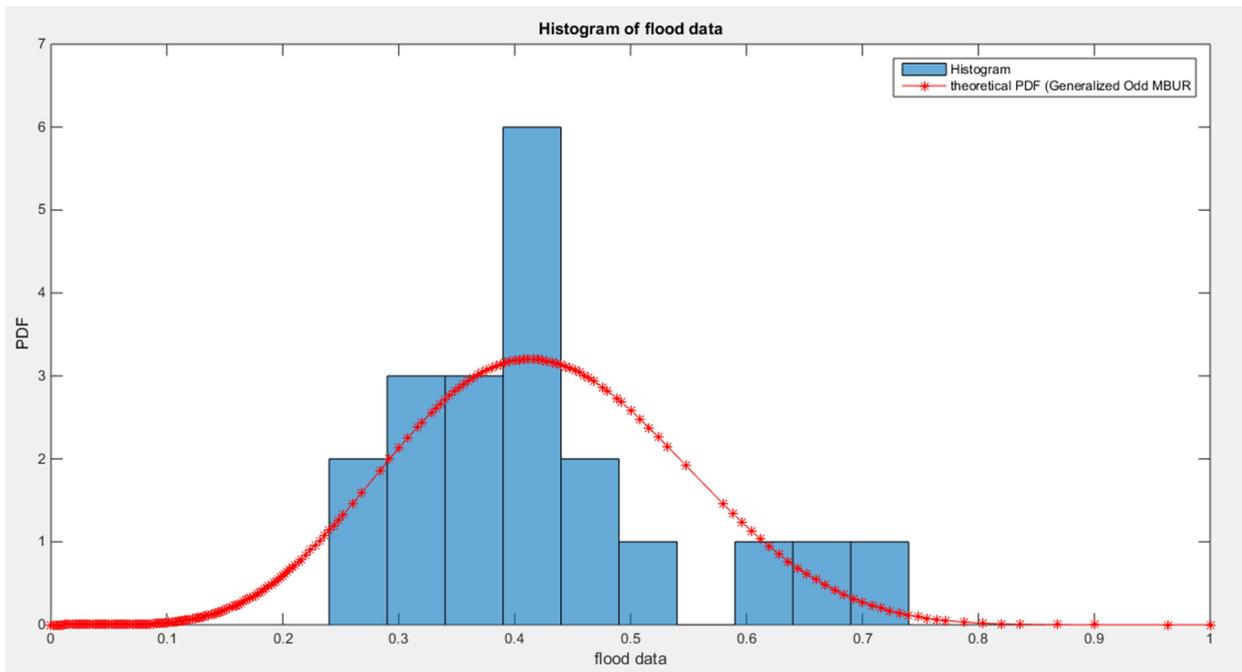

Fig 52 shows the histogram of the flood data and the fitted Generalized Odd MBUR distribution (second version)

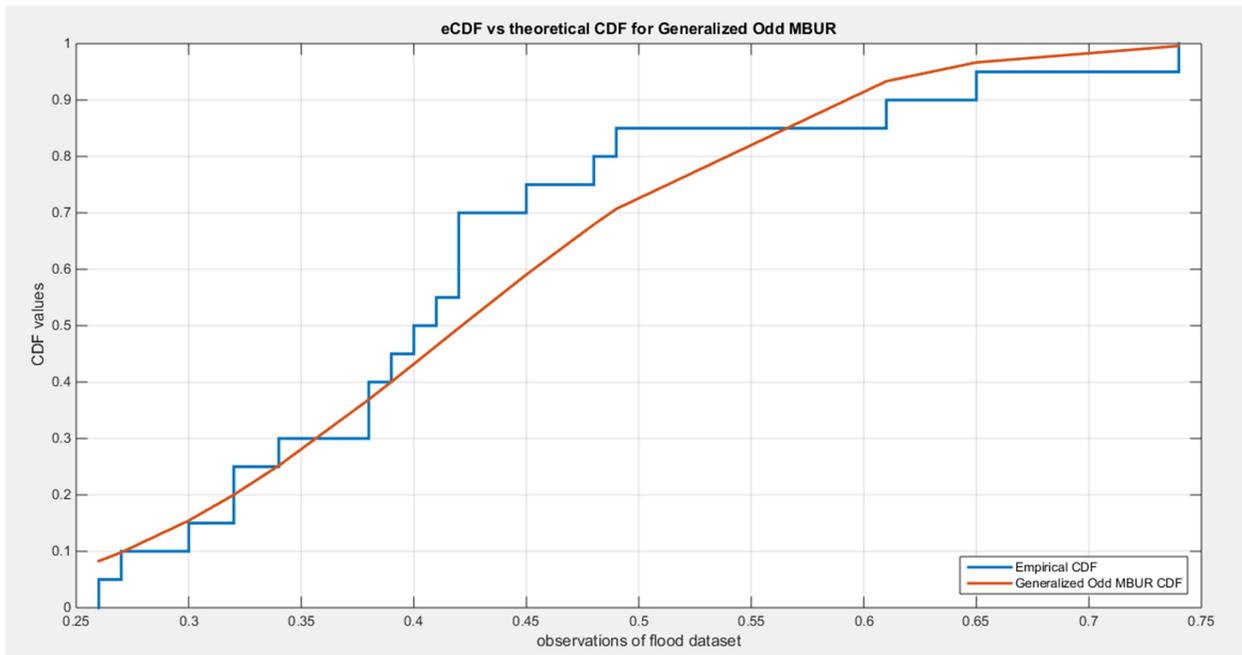

Fig 53 shows the empirical and the theoretical CDF after fitting the Generalized Odd MBUR distribution (second version).

# Section 4: Conclusion

The addition of new parameter to the previously studied MBUR distribution enhances the capability of BMUR to fit the data. Two versions for generalizing the MBUR were discussed by the author. Both have a non-explicit closed form of the CDF, but they can be expressed using special function. Subsequently, the quantile functions for both versions are not expressed in closed form. Also the $r^{th}$ moments of both versions are expressed with the aid of special function. This is considered as a limitation for being used in applications like median based quantile regression or mean based regression in generalized linear model. The advantage of adding this new parameter is that it helps BMUR to fit near symmetric data. Moreover, the two versions of this Generalized Odd MBUR exhibit a new shape for the hazard rate in the form of oscillating patter at the end of the distribution before approaching infinity and at different values of the random variable depending on the level of the alpha and the n parameter.

# Future work:

Using Bayesian analysis can enhance the analysis of the data.

**Declarations:**
**Ethics approval and consent to participate**
Not applicable.
**Consent for publication**
Not applicable
**Availability of data and material**
Not applicable. Data sharing not applicable to this article as no datasets were generated or analyzed during the current study.
**Competing interests**
The author declares no competing interests of any type.
**Funding**
No funding resource. No funding roles in the design of the study and collection, analysis, and interpretation of data and in writing the manuscript are declared
**Authors' contribution**
AI carried the conceptualization by formulating the goals, aims of the research article, formal analysis by applying the statistical, mathematical and computational techniques to synthesize and analyze the hypothetical data, carried the methodology by creating the model, software programming and implementation, supervision, writing, drafting, editing, preparation, and creation of the presenting work.
**Acknowledgement**
Not applicable

# References:

1. Johnson NL. Systems of Frequency Curves Generated by Methods of Translation. Biometrika. 1949;36(1/2):149–76. https://www.jstor.org/stable/2332539

2. Eugene N, Lee C, Famoye F. Beta-Normal Distribution And Its Applications. Communications in Statistics - Theory and Methods. 2002;31(4):497–512.: http://www.tandfonline.com/doi/abs/10.1081/STA-120003130

3. Gündüz S, Korkmaz MÇ. A New Unit Distribution Based On The Unbounded Johnson Distribution Rule: The Unit Johnson SU Distribution. Pak.j.stat.oper.res. 2020;471–90: https://pjsor.com/pjsor/article/view/3421

4. Topp CW, Leone FC. A Family of J-Shaped Frequency Functions. Journal of the American Statistical Association. 1955;50(269):209–19.: http://www.tandfonline.com/doi/abs/10.1080/01621459.1955.10501259

5. Consul PC, Jain GC. On the log-gamma distribution and its properties. Statistische Hefte. 1971;12(2):100–6. http://link.springer.com/10.1007/BF02922944

6. Grassia A. On A Family Of Distributions With Argument Between 0 And 1 Obtained By Transformation Of The Gamma And Derived Compound Distributions. Australian Journal of Statistics. 1977;19(2):108–14. https://onlinelibrary.wiley.com/doi/10.1111/j.1467-842X.1977.tb01277.x

7. Mazucheli J, Menezes AFB, Dey S. Improved maximum-likelihood estimators for the parameters of the unit-gamma distribution. Communications in Statistics - Theory and Methods. 2018;47(15):3767–78. https://www.tandfonline.com/doi/full/10.1080/03610926.2017.1361993

8. Tadikamalla PR. On a family of distributions obtained by the transformation of the gamma distribution. Journal of Statistical Computation and Simulation. 1981;13(3–4):209–14. http://www.tandfonline.com/doi/abs/10.1080/00949658108810497

9. Tadikamalla PR, Johnson NL. Systems of frequency curves generated by transformations of logistic variables. Biometrika. 1982;69(2):461–5. https://academic.oup.com/biomet/article-lookup/doi/10.1093/biomet/69.2.461

10. Kumaraswamy P. A generalized probability density function for double-bounded random processes. Journal of Hydrology. 1980;46(1–2):79–88. https://linkinghub.elsevier.com/retrieve/pii/0022169480900360

11. Modi K, Gill V. Unit Burr-III distribution with application. Journal of Statistics and Management Systems. 2020 ;23(3):579–92. https://www.tandfonline.com/doi/full/10.1080/09720510.2019.1646503

12. Haq MAU, Hashmi S, Aidi K, Ramos PL, Louzada F. Unit Modified Burr-III Distribution: Estimation, Characterizations and Validation Test. Ann Data Sci . 2023;10(2):415–40. https://link.springer.com/10.1007/s40745-020-00298-6

13. Korkmaz MÇ, Chesneau C. On the unit Burr-XII distribution with the quantile regression modeling and applications. Comp Appl Math  2021;40(1):29. https://link.springer.com/10.1007/s40314-021-01418-5

14. Mazucheli J, Maringa AF, Dey S. Unit-Gompertz Distribution with Applications. Statistica. 2019;Vol 79:25-43 Pages. https://rivista-statistica.unibo.it/article/view/8497

15. Mazucheli J, Menezes AFB, Chakraborty S. On the one parameter unit-Lindley distribution and its associated regression model for proportion data. Journal of Applied Statistics. 2019;46(4):700–14. https://www.tandfonline.com/doi/full/10.1080/02664763.2018.1511774

16. Mazucheli J, Menezes AFB, Fernandes LB, De Oliveira RP, Ghitany ME. The unit-Weibull distribution as an alternative to the Kumaraswamy distribution for the modeling of quantiles conditional on covariates. Journal of Applied Statistics. 2020;47(6):954–74. https://www.tandfonline.com/doi/full/10.1080/02664763.2019.1657813

17. Mazucheli ,J., Menezes, A.F., Dey,S., The Unit-Birnbaum-Saunders distribution with applications. Chelian Journal of Statisitcs. 2018;9:47–57.

18. Maya R, Jodrá P, Irshad MR, Krishna A. The unit Muth distribution: statistical properties and applications. 2024;73(4):1843–66. https://link.springer.com/10.1007/s11587-022-00703-7

19. Dumonceaux R, Antle CE. Discrimination Between the Log-Normal and the Weibull Distributions.Technometrics.1973;15(4):923–6. http://www.tandfonline.com/doi/abs/10.1080/00401706.1973.10489124